\documentclass[aps, prd, twocolumn, lengthcheck, superscriptaddress, 
nofootinbib]{revtex4-1}

\usepackage{amsfonts}
\usepackage{amsmath}
\usepackage{amssymb}
\usepackage{xcolor}
\usepackage{graphicx}
\usepackage{subfigure}
\usepackage{slashed}
\usepackage{hyperref}
\usepackage{amsthm}
\usepackage{mathrsfs}
\usepackage{color}
\usepackage{epsfig}
\usepackage{epstopdf}
\usepackage{bm}
\usepackage[T1]{fontenc}
\usepackage{ae,aecompl}	
\usepackage[toc,page]{appendix}
\usepackage{soul}
\usepackage[utf8]{inputenc}
\usepackage{multirow}
\usepackage{verbatim}
\usepackage{threeparttable}

\def\be{\begin{eqnarray}}
	\def\ee{\end{eqnarray}}

\allowdisplaybreaks{}
\begin{document}
	
\newcommand*{\ssp}{\scriptscriptstyle}
\newcommand*{\gm}{\gamma}
\newcommand*{\nm}{\nonumber}

\title{Chiral-scale effective field theory for dense and thermal systems 
}
\author{Jia-Ying Xiong}
\affiliation{School of Fundamental Physics and Mathematical Sciences, Hangzhou Institute for Advanced Study, UCAS, Hangzhou, 310024, China}
\affiliation{Institute of Theoretical Physics, Chinese Academy of Sciences, Beijing, 100190, China}
\affiliation{University of Chinese Academy of Sciences, Beijing, 100049, China}
\affiliation{School of Frontier Sciences, Nanjing University, Suzhou, 215163, China}

\author{Yao Ma}
\email{mayao@nju.edu.cn}
\affiliation{School of Frontier Sciences, Nanjing University, Suzhou, 215163, China}

\author{Bing-Kai Sheng}
\email{bingkai.sheng@ucas.ac.cn}
\affiliation{School of Fundamental Physics and Mathematical Sciences, Hangzhou Institute for Advanced Study, UCAS, Hangzhou, 310024, China}
	
\author{Yong-Liang Ma}
\email{ylma@nju.edu.cn}
\affiliation{School of Frontier Sciences, Nanjing University, Suzhou, 215163, China}
\affiliation{International Center for Theoretical Physics Asia-Pacific (ICTP-AP) , UCAS, Beijing, 100190, China}
\date{\today}	

\begin{abstract}
We established a new power counting scheme, chiral-scale density counting (CSDC) rules, for the application of the chiral-scale effective field theory to nuclear matter at finite densities and temperatures. Within this framework, the free fermion gas is at the leading order, while one-boson-exchange interactions appear at the next-to-leading order, and the multimeson couplings are at higher orders. Then, we applied the CSDC rules to study the nuclear matter properties, and estimated the valid regions of the CSDC rules. It was found that the zero temperature symmetric nuclear matter properties around saturation density and the critical temperature of liquid-gas phase transition can be captured by an appropriate choice of CSDC orders, and the results beyond these regions are align with the chiral nuclear force. Moreover, the evolution of scale symmetry was found to be consistent with previous studies. The results of this work indicate that the quantum corrections may be crucial in the studies of nuclear matter in a wide density region.
\end{abstract}

\maketitle

\section{Introduction}\label{sec:intro}

Investigating the thermodynamical properties of strongly interacting matter is one of the most important topics in high energy nuclear physics~\cite{Gross:2022hyw}.
On the one hand, quantum chromodynamics (QCD) is nonperturbative at low energies, so that it cannot be applied to hadron phenomenon analysis straightforwardly.
On the other hand, it is inferred that the extreme environments with high temperature and/or density exist in the early Universe~\cite{Kolb:1990vq}, ultrarelativistic heavy-ion collisions (HIC) at RHIC and LHC~\cite{Gyulassy:2004zy,ATLAS:2010isq}, and cores of massive neutron stars (NSs)~\cite{Lattimer:2015eaa,doi:10.1142/11072}.
The energy scale of these extreme environments is near that of QCD, i.e., $\Lambda_{\mathrm{QCD}}\approx200~\text{MeV}$ and hence they are instrumental in probing the nonperturbative properties of QCD.
So, the studies on the thermodynamic properties of nuclear matter (NM) and/or quark matter in the extreme environments both experimentally and theoretically can deepen our understanding of QCD in nonperturbative regions and the evolution of Universe~\cite{1990eaun.book.....K,Abe:2020sqb}.
Extensive studies have been implemented in recent decades for the chiral phase transition~\cite{Pisarski:1983ms,Brown:1990ev,Pelissetto:2013hqa,Bonati:2015bha,Kohyama:2015hix,Bonati:2018nut,HotQCD:2019xnw,HotQCD:2018pds,Buballa:2018hux,Borsanyi:2020fev,Cuteri:2021ikv,Kotov:2021rah,Lahiri:2021lrk,Pannullo:2024sov} and the confinement-deconfinement phase transition~\cite{Kajantie:1981wh,Yaffe:1982qf,Kogut:1982rt,Fukushima:2003fw,Kaczmarek:2005ui,Ratti:2005jh,Andreev:2009zk,Hell:2009by,Sakai:2010rp,Dexheimer:2009hi,Ruggieri:2012ny,Bellwied:2013cta,Fukushima:2017csk,Mattos:2021tmz}.

As a fundamental quantity in thermodynamics, the equation of state (EOS) of QCD matter which is usually expressed by the pressure $p$ as a function of temperature and chemical potentials can bring us information about the properties of strongly interacting matter.
In recent years, many works have been performed for the investigation of phase structures with the EOSs~\cite{Gupta:2011wh,HotQCD:2017qwq,Borsanyi:2020fev,Fukushima:2008wg,Ratti:2005jh,Costa:2010zw,Sasaki:2011cj,Ferreira:2013oda,Schaefer:2009ui,Skokov:2010uh,Qin:2010nq,Gao:2015kea,Fischer:2014ata,Fu:2019hdw,Stephanov:2008qz,Stephanov:2011pb,Shao:2017yzv,Shao:2019hen,Luo:2015ewa,Luo:2015doi,Luo:2017faz,STAR:2020tga,Song:2010mg,Song:2010aq,Deb:2016myz,Leonhardt:2019fua}. Theoretically, EOSs of strongly interacting matter can be calculated by the lattice QCD (LQCD) simulations as {\it ab initio} methods~\cite{Panero:2009tv,Borsanyi:2012cr,HotQCD:2014kol,Borsanyi:2013bia,Bollweg:2022rps}.
In Ref.~\cite{Panero:2009tv}, the thermodynamic quantities such as pressure, energy density, entropy density are systematically evaluated in pure-$\mathrm{SU(N_{\mathrm{c}})}$ gauge sector at finite temperature.
When the quark degree of freedom is involved, the QCD EOS at vanishing baryon chemical potential is still well controlled in LQCD simulations, e.g., the speed of sound~\cite{HotQCD:2014kol,Borsanyi:2013bia}.
At finite baryon density or baryon chemical potential $\mu_{\mathrm{B}}$, the fermion sign problem in LQCD brings barriers on the way to a high baryon density region. As a complement to the first-principle calculation, the effective models of QCD play a crucial role in the study of EOSs of strongly interacting matter. For example, the hadron resonance gas (HRG) model~\cite{Venugopalan:1992hy,Huovinen:2017ogf,Kadam:2019peo,Biswas:2024xxh}, the field correlator method (FCM)~\cite{Khaidukov:2018lor,Khaidukov:2019icg}, and the quasiparticle model~\cite{Mykhaylova:2020pfk}.
And well-known constituent quark models such as Nambu\textendash Jona-Lasinio (NJL)-type models~\cite{Ghosh:2006qh,Deb:2016myz,Motta:2020cbr,Marty:2013ita,Saha:2017xjq,Xue:2021ldz,Pasqualotto:2023hho} and the quark-meson model~\cite{Schaefer:2009ui,Abhishek:2017pkp,Rai:2023wtz,Yang:2023duo,Zhang:2020zrv} have also been utilized in these studies.

Despite extensive prior researches of the EOSs, which mainly focus on high density or/and high temperature, the properties of NM at finite temperature and density, especially at tens of MeV of temperature and several times saturation density $n_0$, are still shrouded in deep mystery.
Traditionally, the properties of NM at this regime are described based on several parametrizations, for example, Walecka-type models~\cite{Walecka:1974qa,Serot:1997xg} and chiral nucleon force (\(\chi\)NF) models with in-medium chiral perturbation theory~\cite{Epelbaum:2008ga}.
However, these approaches all have their limitations: Walecka-type models lack the consideration of the chiral symmetry pattern of QCD---a significant factor should be considered in the low energy strong interaction, and the introduction of meson exchange in \(\chi\)NF models is not well established for high-density systems.
On the other hand, as a systematic low-energy field theory of QCD, the original chiral effective field theory (\(\chi\)EFT)---referred to as ``standard $\chi$EFT" (S$\chi$EFT for short)---takes only the nucleon (matter field) and the pion (Nambu-Goldstone bosons associated with the chiral symmetry breaking) as the relevant degrees of freedom in nuclear dynamics, and effective operators are organized by the chiral (power) counting series.
The S$\chi$EFT has been widely applied in hadron and nuclear physics, see, e.g., Refs.~\cite{Gasser:1983yg,Gasser:1984gg,Gasser:1987rb,Machleidt:2011zz}. Being an EFT with pions only (apart from the nucleons), the theory is defined with the cutoff set typically at $\Lambda \sim (400-500)$~MeV, therefore describes the pion and nucleon experimental data available to $\sim 350$~MeV and the nuclear matter up to the saturation density $n_0 \approx 0.16$~fm$^{-3}$ which corresponds to $k_F=1.68~$fm$^{-1} \sim 330$~MeV. The resonances above this energy scale, such as the vector mesons $\rho$ and $\omega$ and scalar meson $\sigma$ which are critical to nuclear force with energy change $\sim 800$~MeV corresponding to nuclear force and nuclear matter properties at the center of massive stars $\sim 10 n_0$, are integrated out from the S\(\chi\)EFT Lagrangian. It is well known that, in order to describe the dense nuclear matter properties, the repulsive and attractive forces between nucleons captured by the vector and scalar meson exchanges are indispensable.
Therefore, the S$\chi$EFT should be extended to include these resonances.

Regarding the above facts, it is essential to study the properties of dense NM based on a reliable effective theory of QCD. 
In this work, we propose to study the EOS of NM by applying an EFT based on the nonlinear realization of chiral symmetry and scale symmetry of QCD at low energies---\(bs\)HLS~\cite{Sasaki:2011ff,Lee:2015qsa,Paeng:2015noa,Li:2016uzn}, where $b$ stands for baryon, $s$ for scalar (dilaton) and HLS for hidden local symmetry.
The \(bs\)HLS is based on the S\(\chi\)EFT and the lowest-lying vector mesons $\rho$ and $\omega$ are included through the hidden local symmetry (HLS) approach~\cite{Bando:1984ej,Bando:1987br, Harada:2003jx}, and the lightest scalar meson $\sigma$ is identified as a Nambu–Goldstone mode, ``dilaton," of scale symmetry~\cite{Crewther:2013vea, Cata:2018wzl}.
Different from the Walecka-type models, the $bs$HLS is anchored on the fundamental symmetry of QCD, the chiral symmetry, scale symmetry and local flavor symmetry, and has a self-consistent power counting mechanism.
Using the leading order scale symmetry (LOSS) approximation of $bs$HLS, where the trace anomaly of QCD is only encoded in the dilaton potential, some interesting and unique features in the properties of dense nuclear matter have been found:
(I). A pseudoconformal structure of compact star matter that the trace of energy momentum tensor does not vanish but the sound velocity saturates the conformal limit was found~\cite{Paeng:2017qvp,Ma:2018jze,Ma:2018xjw}; (II) A genuine peak of sound velocity at intermediate densities \(\sim (1.5-2.5)\ n_0\) with only a pure hadron phase is yielded~\cite{Zhang:2024sju}. The Walecka-type models or linearization of low energy QCD never show such results;
(III). A supermassive neutron star with \(M_{\rm max}\sim 2.85\ M_{\odot}\), while satisfying the constraints from the GW170817 event~\cite{LIGOScientific:2018cki,LIGOScientific:2018hze}, see Ref.~\cite{Zhang:2024iye}.

So far, our works are limited to zero temperature without any thermal properties~\cite{Zhang:2024sju,Zhang:2024iye}, we intend to extend the two-flavor $bs$HLS to both finite density and temperature system in this work.
We set up a new counting rule in medium beyond the chiral-scale counting rule in matter-free space---the chiral-scale density order counting (CSDC) rules, which is applicable in the cores of massive stars up to $n \lesssim 10 n_0$, or equivalently, Fermi momentum $k_F \approx 700$~MeV, and can be consistently extended to finite temperature systems (roughly $T \lesssim 100$~MeV) via perturbation approximations.
With the CSDC rules, the Lagrangian/Hamiltonian of the $bs$HLS can be expanded by a characteristic momentum \(k_{\rm{c}}\). It is found that:
The leading order (LO) contribution to thermal quantities is the free Fermi fluid term;
The next-to-leading order (NLO) contribution is the OBE part; The next-to-next-to-leading order (N$^2$LO) contribution is the pion derivative coupling and terms involved with multimeson interactions;
The next-to-next-to-next-to-leading order (N$^3$LO) contribution is the four-meson self-coupling terms, and so on.
The complexity of calculating certain order terms is equivalent to the corresponding number of fermion loops in the Feynman diagrams, but can be simplified with the help of relativistic mean field (RMF) approximation.

Within current framework, it is found that NM properties around saturation density and critical temperature of gas-liquid phase transition (GLPT) can be reproduced by the appropriate choice of the CSDC order and corresponding parameters under RMF approximation, and \(\mathcal{O}(k_{\rm{c}}^{12})\) is good enough to capture all these physics.
Then, scale symmetry is found to restore at low densities, but rebreaks at high densities due to the unique coupling constraints from the scale symmetry. This will lead a kink behavior of the sound velocity at intermediate densities.
Such a behavior is consistent with what we found in Ref.~\cite{Zhang:2024sju} at zero temperature.

The paper is organized as follows:
In Sec.~\ref{sec:TF}, the CSDC rules of chiral-scale EFT for NM, is first established at finite density and zero temperature with relativistic Hartree-Fock (RHF), then consistently extended to finite temperature regions;
In Sec.~\ref{sec:res}, the NM properties around saturation density at zero temperature are discussed with different CSDC orders, and the scale symmetry behavior at different densities and temperatures will also be presented;
The summary and outlook is given in the last section.

\section{Theoretical framework}
\label{sec:TF}

In this section, we introduce the methodology of our framework for calculating thermal quantities of NM.
We start with the discussion of \(bs\)HLS and its CSDC rule for zero-temperature NM.
Then, we discuss the mean-field method on handling the dense NM and the fermion loop expansion. Finally, we establish the CSDC rules, and consistently extend the framework to finite temperature.

\subsection{$bs$HLS and CSDC rule at zero temperature}

The chiral-scale EFT was formulated in three-flavor QCD such that the lightest scalar meson $\sigma$, which has a similar mass as the kaon, can be regarded as the Nambu-Goldstone boson~\cite{Crewther:2013vea}.
Here, we simply consider its two-flavor reduction and suppose that its validity is intact. Following the procedure discussed in detail in Ref.~\cite{Li:2016uzn}, we write the $bs$HLS Lagrangian at the leading order of chiral-scale counting (to be specified below) as
\begin{equation}
\mathcal{L}=\mathcal{L}_M+\mathcal{L}_B ,
\label{eq:LbsHLS}
\end{equation}
where \(\mathcal{L}_M\) and \(\mathcal{L}_B\) are, respectively, the meson and the baryon parts.

The meson part  \(\mathcal{L}_M\) is written as
\be
\label{eq:LM}
\mathcal{L}_M & = & \ f^2_\pi\Phi^2{\rm Tr}\left(\hat{\alpha}^\mu_\perp\hat{\alpha}_{\mu\perp}\right) +\frac{m^2_\rho}{g^2_\rho}\Phi^2{\rm Tr}\left(\hat{\alpha}^\mu_\parallel\hat{\alpha}_{\mu\parallel}\right) \nonumber \\
& &{} +\frac{1}{2}\left(\frac{m^2_\omega}{g^2_\omega}-\frac{m^2_\rho}{g^2_\rho}\right)\Phi^2{\rm Tr}\left(\hat{\alpha}^\mu_\parallel\right){\rm Tr}\left(\hat{\alpha}_{\mu\parallel}\right) \nonumber\\
& &{} -\frac{1}{2g^2_\rho}{\rm Tr}\left(V_{\mu\nu}V^{\mu\nu}\right)-\frac{1}{2g^2_0}{\rm Tr}\left(V_{\mu\nu}\right){\rm Tr}\left(V^{\mu\nu}\right) \nonumber\\
& &{} +\frac{1}{2}\partial_\mu\chi\partial^\mu\chi+\frac{f^2_\pi}{4}\Phi^2{\rm Tr}\left(\mathcal{M}U^\dagger+U\mathcal{M}^\dagger\right) \nonumber\\
& &{} +h_5\Phi^4+h_6\Phi^{4+\beta'},
\ee
where $V_{\mu \nu}=\partial_\mu V_\nu-\partial_\nu V_\mu-i\left[V_\mu, V_\nu\right]$, and $V_\mu=\frac{1}{2}\left(g_\omega \omega_\mu+g_\rho \rho_\mu^a \tau^a\right)$ with $\tau^a$ being the Pauli matrices. $m_\rho$, $m_\omega$ and \(\mathcal{M}=m_{\pi}^2 I_{2\times2}\) are the masses of $\rho$, $\omega$, and \(\pi\) mesons, respectively.
Scalar meson \(\sigma\) is introduced as a nonlinear representation---dilaton field $\chi=f_\chi\Phi=f_\chi \exp \left(\sigma / f_\chi\right)$, and \(\beta'\) accounts for the anomalous dimension of gluon field operators, representing the deviation from the conjectured IR fixed point (IRFP)~\cite{Crewther:2013vea}. Pions are introduced as a nonlinear field $\xi=\sqrt{U}=e^{i \frac{\pi}{f_{\pi}}}$, where $\pi=\pi^a \tau^a/2$.
The Maurer-Cartan 1-forms $\hat{\alpha}_{\perp}^\mu$ and $\hat{\alpha}_{\|}^\mu$ are defined as
\begin{equation}
	\hat{\alpha}_{\perp,\| }^\mu=\frac{1}{2 i}\left(D^\mu \xi \cdot \xi^{\dagger}\mp D^\mu \xi^{\dagger} \cdot \xi\right),
\end{equation}
where the covariant derivative is defined as
\begin{equation}
	D_\mu \xi
    =\left[\partial_\mu-i\frac{1}{2}\left(g_\omega\omega_\mu+g_\rho\rho_\mu^a\tau^a\right)\right] \xi.
\end{equation}
With an appropriate choice of the signs of the parameters such that the dilaton potential has spontaneous symmetry breaking, $h_5$ and $h_6$ are constrained by the saddle point equation and dilaton mass through 
\be
\label{eq:sp}
& & 4 h_5+\left(4+\beta^{\prime}\right) h_6+2 m_\pi^2 f_\pi^2=0 , \nonumber\\
& & 12 h_5+\left(4+\beta^{\prime}\right)\left(3+\beta^{\prime}\right) h_6+2 m_\pi^2 f_\pi^2=-m_\sigma^2 f_\chi^2 .
\ee
In Lagrangian~\eqref{eq:LM}, the trace anomaly effect is only encoded in the dilaton potential and other parts are scale invariant. This is referred to as leading-order scale symmetry (LOSS) approximation~\cite{Ma:2019ery}.

The baryonic part Lagrangian \(\mathcal{L}_B\) is written as
\be
\label{eq:LB}
\mathcal{L}_B & = & \bar{N} i \gamma_\mu D^\mu N-m_N \Phi \bar{N} N \nonumber\\
& &{} +\left[g_A C_A+g_A\left(1-C_A \right)\Phi^{\beta^{\prime}}\right] \bar{N} \hat{\alpha}_{\perp}^\mu \gamma_\mu \gamma_5 N \nonumber\\
& &{} +\left[g_{V_\rho} C_{V_\rho}+g_{V_\rho}\left(1-C_{V_\rho}\right) \Phi^{\beta^{\prime}}\right] \bar{N} \hat{\alpha}_{\|}^\mu \gamma_\mu N \nonumber\\
& &{} +\frac{1}{2}\left[g_{V_0} C_{V_0}+g_{V_0}\left(1-C_{V_0}\right) \Phi^{\beta^{\prime}}\right] \operatorname{Tr}\left[\hat{\alpha}_{\|}^\mu\right] \bar{N} \gamma_\mu N ,
\nonumber\\
\ee
where \(N\) is the nucleon doublet and $D^\mu=\partial^\mu+\frac{1}{2}g_\omega\omega^\mu+\frac{1}{2}g_\rho\rho^{\mu, a}\tau^a$. Here, we considered the corrections to the LOSS that does not contribute to the anomaly match in matter-free space but are found to be significant for understanding the weak decay of heavy nuclei~\cite{Ma:2020tsj} and emergence of scale symmetry in nuclear medium~\cite{Zhang:2024sju,Zhang:2024iye}.

For the later convenience, we introduce the combinations of the parameters
\be
& & g_{\omega NN}=\frac{1}{2}\left(g_{V_{\rho}} C_{V_{\rho}}+g_{V_0} C_{V_0}-1\right) g_\omega ,\nonumber\\
& & g_{\rho NN}=\frac{1}{2}\left(g_{V_{\rho}} C_{V_{\rho}}-1\right) g_\rho ,\nonumber\\
& & g_{\pi N N}=-\frac{\left(g_A C_A\right)}{2f_\pi},\nonumber\\
& & g^{SSB}_{\omega NN}=\frac{1}{2}\left[g_{V_\rho}\left(1-C_{V_\rho}\right)+g_{V_0}\left(1-C_{V_0}\right)\right]g_\omega ,\nonumber\\
& & g^{SSB}_{\rho NN}=\frac{1}{2}\left[g_{V_\rho}\left(1-C_{V_\rho}\right)\right]g_\rho ,\nonumber\\
& & g^{SSB}_{\pi N N}=-\frac{\left[g_A\left(1-C_A\right)\right]}{2f_\pi} .
\ee
The Lagrangian~\eqref{eq:LB} is the \(\mathcal{O}(p)\) level Lagrangian, together with the meson part~\eqref{eq:LM}, forming the leading order Lagrangian of the $bs$HLS for NN interactions.

Now, let us discuss the chiral-scale power counting rule briefly.
As explicitly discussed in Ref.~\cite{Li:2016uzn}, the chiral-scale power counting rule is an extension of that of \(\chi\)EFT~\cite{Bando:1987br,Harada:2003jx}, because the $bs$HLS is based on the nonlinear realization of chiral symmetry and the HLS.
Similar to \(\chi\)EFT, all the derivatives acting on the Nambu-Goldstone bosons, here pion and dilaton, are order $\mathcal{O}(p)$, and the quark masses are counted as $\mathcal{O}(p^2)$.
For the baryon fields, since they are nonrelativistic, the combination \((\slashed{p}-m_B)\) is of order \(\mathcal{O}(p)\), while the corresponding \(\slashed{p}\) and \(m_B\) terms are of order \(\mathcal{O}(1)\).
For the operators involving vector meson fields, they have the counting order $\hat{\alpha}_{\parallel \mu} \sim \frac{1}{g}V_{\mu\nu} \sim \mathcal{O}(p)$.
In addition to these counting rules which are the same as those in \(\chi\)EFT and HLS, a new rule involves due to the deviation from the IRFP $\Delta\alpha_s=\alpha_s - \alpha_{IR}$.
From the $\sigma$-to-vacuum matrix element of the divergence of the dilaton current, one concludes $\Delta\alpha_s \sim \mathcal{O}(p^2)$, for the fact that it is proportional to the dilaton mass square.
Considering the scale compensator coupled to nucleon mass, the expansion \(\Phi-1\) accounting for the deviation of the nucleon mass in medium is treated as \(\mathcal{O}(p)\). The couplings $g_{\sigma^nNN}    (n=1,2,\cdots)$  between \(\sigma^n\) and nucleon stemming from $(\Phi-1)\bar{N}N$ must have the same order as the couplings between the vector mesons and nucleon, due to the cancellation mechanism for the saturation of NM. As the result, an additional assumption we would like to made is that the coupling $g_{\sigma^nNN}$, which can be expressed as $m_N$ and $f_\chi$, is at $\mathcal{O}(p)$. Then, the power-counting scheme is consistent for both baryonic and mesonic parts. These counting rules enable nonlinear \(\chi\)EFT to outperform linear ones in low-energy process calculations, as the Lagrangians and Feynman diagrams are systematically organized to capture the physics order by order.

\subsection{Relativistic Hartree-Fock method and fermion loop expansion}

Consider a system with baryons at finite density and zero temperature.
Calculating the EOS of such a system is equivalent to solving the corresponding Dirac equations of baryons in momentum space~\cite{Bouyssy:1987sh}
\begin{equation}
\label{eq:Dirac}
\left(\slashed{p}-m_B-\Sigma(\slashed{p})\right)\psi_B=0 ,
\end{equation}
where \(\psi_B\) is the fermion field of baryons with $m_B$ as its mass in the matter-free space, and \(\Sigma=\gamma^{\mu}\Sigma_{\mu}+\Sigma_0\) is the self-energy of baryons, which is determined by the interactions between baryons and mesons.
Then, the effective momentum and mass of a baryon in the medium are given by
\begin{equation}
\slashed{p}^*=\slashed{p}-\slashed{\Sigma},\ m^*_B=m_B+\Sigma_0 .
\end{equation}
Note that, in this work, for simplicity, we setup the framework in a Lorentz invariant form. The extension to $\rm SO(3)$ invariant form is straightforward. Then, in momentum space, the Dirac equation can be written as
\begin{equation}
	\left[\mathbf{\gamma} \cdot \mathbf{p}^*+m_B^*\right] u(\mathbf{p}, s)=\gamma_0 E^* u(\mathbf{p}, s) ,
\end{equation}
where no-sea approximation is assumed with \(E^{*2}=\mathbf{p}^{*2}+m_B^{*2}\).
The fermion field is expanded as
\begin{equation}
	\label{eq:psiB}
	\psi_B(x)=\sum_{s}\int_0^{\mathbf{k}_{\rm F}} \frac{{\rm d}\mathbf{p}}{(2\pi)^3}\frac{1}{\sqrt{2E^*}}\left[u(\mathbf{p}, s)e^{-ip\cdot x}a(\mathbf{p}, s)\right] ,
\end{equation}
where summation is over the spin of the baryon, \(\mathbf{k}_{\rm F}\) is the Fermi surface of the bare three-momentum in mean-field and perturbation approximations, and $a(\mathbf{p}, s)$ is the annihilation operator of the baryon fields.

On the other hand, the physical vacuum, the ground state of the system, at certain density \(n\) and zero temperature, is defined by the fully occupied Fermi sea made of free fermions, here baryons, and denoted as \(|0\rangle\).
The interaction vacuum can be obtained by the Gell-Mann and Low theorem~\cite{Gell-Mann:1951ooy}: \(|\Omega\rangle\sim U(t,-\infty)|0\rangle\), where \(U(t,-\infty)\) is the time evolution operator, \(T\left\{\exp \left[-i \int_{-\infty}^{t} {\rm d} t' \hat{H}_I(t')\right]\right\}\).
Then, the Hamiltonian of the system can be obtained via
\be
\label{eq:HG}
& & \left\langle\Omega\left|:\hat{\mathcal{H}}:\right|\Omega\right\rangle \\
& & \quad =\left\langle 0\left|T\left\{:\hat{\mathcal{H}}\exp \left[-i \int_{-\infty}^{+\infty} {\rm d} t' \hat{H}_I(t')\right]:\right\}\right|0\right\rangle_c , \nonumber
\ee
\\
where superscript \(c\) denotes the connected Green's function. Note that normal ordered product \(:\hat{\mathcal{H}}:\) is applied to remove the vacuum divergence.

\begin{figure}[htbp]
	\centering
		\centering
		\includegraphics[width=0.45\textwidth]{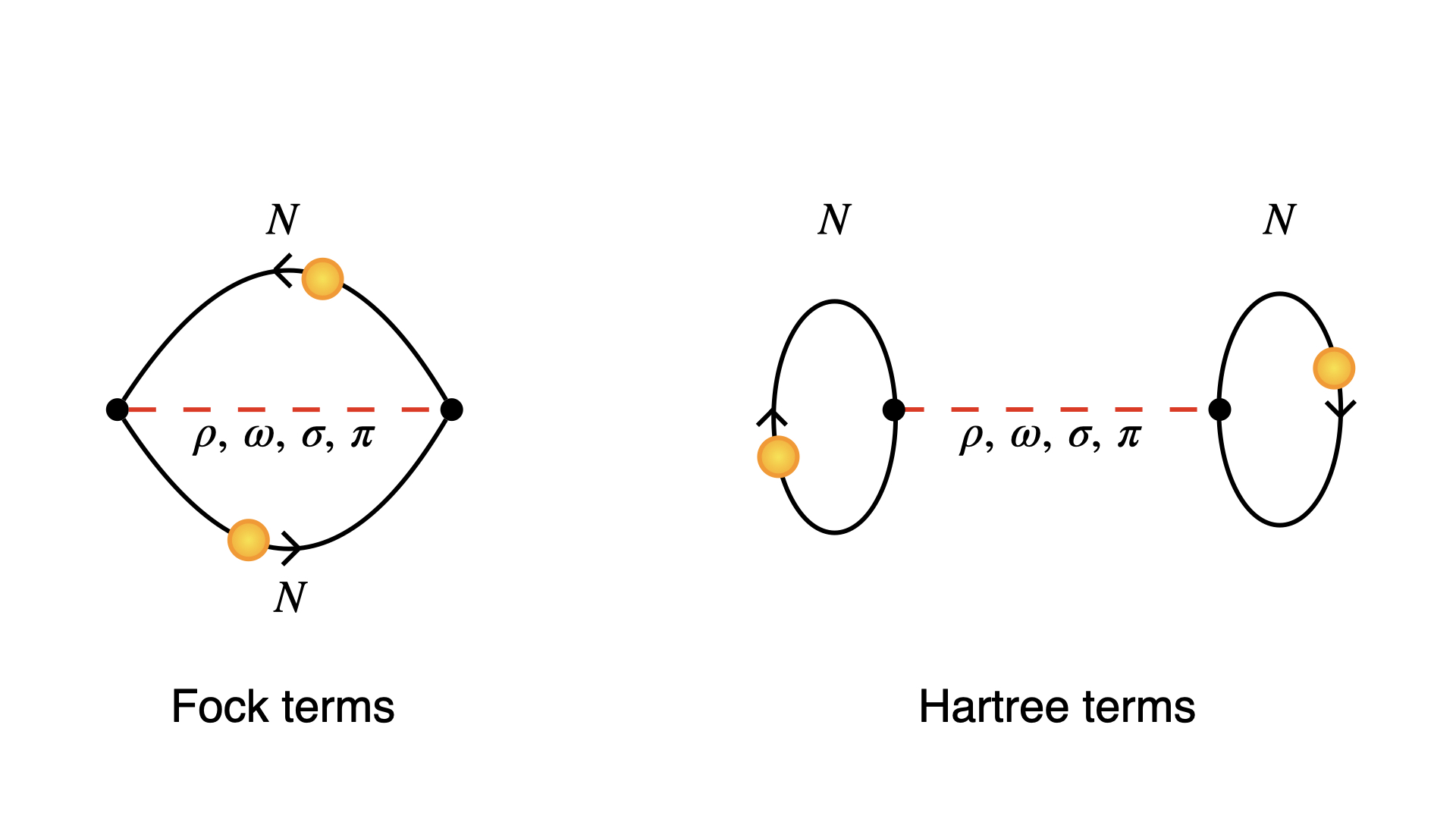}
		\caption{The diagrams for effective four-fermion terms of the RHF method in the OBE model~\eqref{eq:HOBE}.}
	\label{fig:HFOBE}
\end{figure}

To illustrate the calculation process of the EOS of NM, we consider an OBE model as an example.
The Lagrangian of the OBE model is given by
\be
\label{eq:LOBE}
	\mathcal{L}_{\rm OBE} & = & \bar{N}\left[i\slashed{\partial}-m_N-g_{\omega NN}^{\rm OBE}\slashed{\omega}-g_{\rho NN}^{\rm OBE}\slashed{\rho}^i\tau^i \right. \nonumber\\
    & &\left. {}\quad -g_{\sigma NN}^{\rm OBE}\sigma-g_{\pi NN}^{\rm OBE}\slashed{\partial}\pi^i\tau^i\gamma_5\right]N+\mathcal{L}_{\rm meson}, \nonumber\\
\ee
where \(\mathcal{L}_{\rm meson}\) is the free Lagrangian of the corresponding mesons involved in the fermion OBE interactions, and \(N\) is the nucleon isospin doublet.
The EOMs of the mesons, which is treated as the background fields induced by the nucleons, can be solved via the Green's function method with the corresponding fermion currents.
For example, for the $\sigma$ meson, 
\begin{equation}
\label{eq:OGS}
\sigma(x)=\int_V {\rm d}^4 x'D_{\sigma}(x-x')\left(-g_{\sigma NN}^{\rm OBE}\bar{N}(x')N(x')\right) ,
\end{equation}
where \(V\) is the volume of the entire system, and \(D_{\sigma}(x-x')\) is the retarded Green's function of the \(\sigma\) meson field. The other mesons can be solved similarly.
Since the Fock space of the system is only composed of nucleons, the meson fields should be replaced by fermion bilinear operators from a series of solutions like Eq.~\eqref{eq:OGS}.
The Hamiltonian can then be rewritten as follows:
\begin{widetext}
\be
\label{eq:HOBE}
\hat{\mathcal{H}}(x) & = & \bar{N}(x)\left(-i\mathbf{\gamma}\cdot\mathbf{\nabla}+m_N\right)N(x)  +\bar{N}(x)\left(g_{\sigma NN}^{\rm OBE}\int_V {\rm d}^4 x'D_{\sigma}(x-x')\left(-g_{\sigma NN}^{\rm OBE}\bar{N}(x')N(x')\right)\right)N(x)+ \cdots ,
\ee
\end{widetext}
where ``$\cdots$" denotes the other interaction parts, which are omitted for brevity. It can be seen that obtaining the EOS of NM at zero temperature is reduced to solving a series of two- and four-point Green's functions of the fermions in the medium. If Eq.~\eqref{eq:HG} is calculated at the leading order with the form of \(\langle0|:\hat{\mathcal{H}}:|0\rangle\), it will result in the Hartree and Fock diagrams, shown in Fig.~\ref{fig:HFOBE}.

\begin{figure}[htbp]
		\centering
		\includegraphics[width=0.45\textwidth]{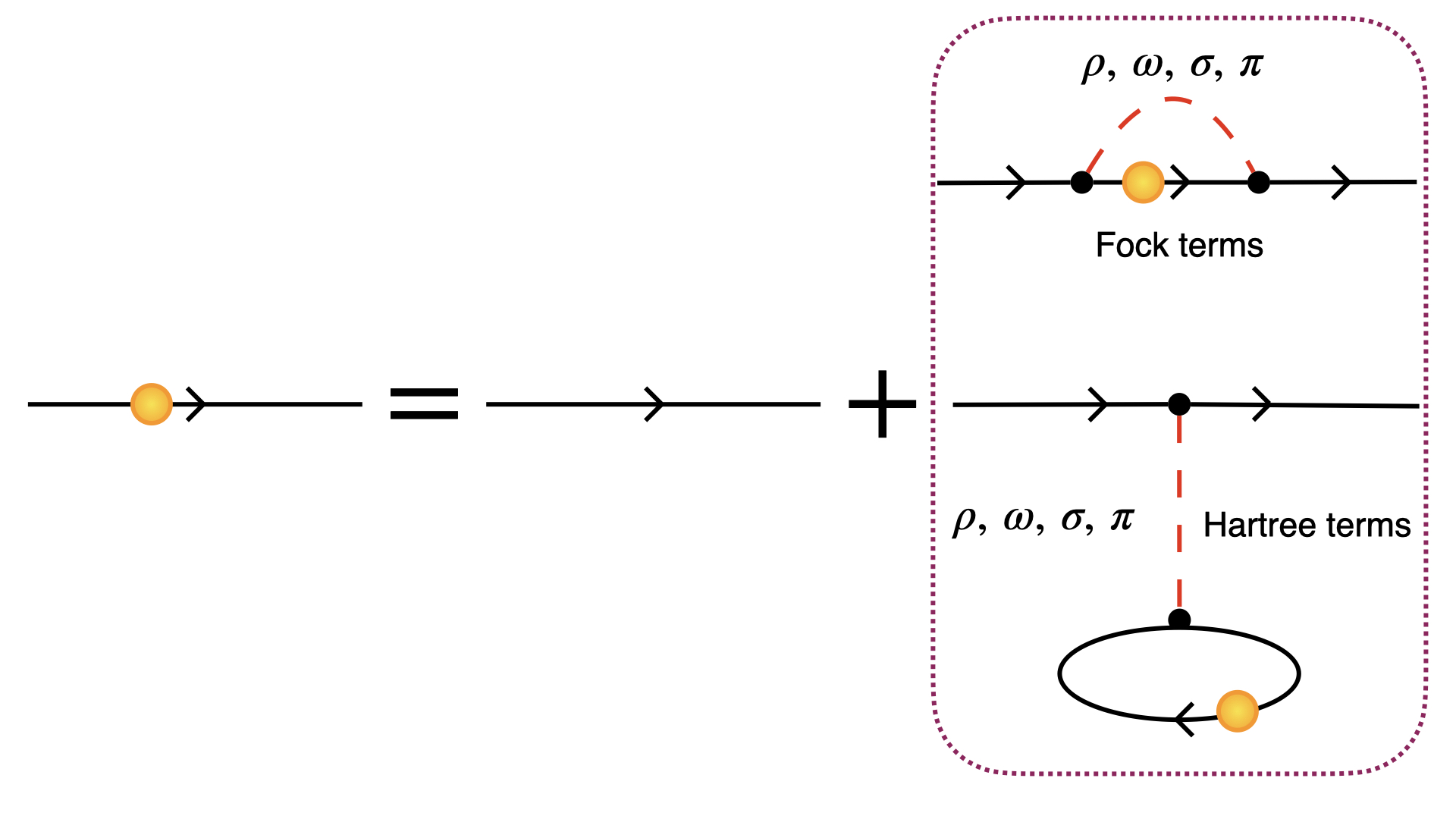}
		\caption{The complete propagators of the nucleons in the medium defined in Eq.~\eqref{eq:Dirac}.}
	\label{fig:InternalE}
\end{figure}
In the RHF method, the internal energy of NM is calculated at the LO level of the Green's function, while the fermion propagators involved are iteratively solved with the integral kernels truncated at the NLO level, as shown in Fig.~\ref{fig:InternalE}. Once the result of Eq.~\eqref{eq:HOBE} is obtained, the EOS of NM can be calculated using thermodynamics.

\subsection{Chiral-scale density order counting rules at zero temperature}
\label{sec:CSDC-0T}

After illustrating RHF approach using the above simple example, we now turn to the $bs$HLS model. Since the dilaton field is not a perturbation field around zero but around \(f_{\chi}\), and it will couple to the mass terms of other fields, which will bring much complexity in the calculation of Green's functions.

To overcome this complexity, we expand the nonlinear terms in the $bs$HLS in a series of physical fields, such as \(\Phi\approx1+\sigma/f_{\chi}+\sigma^2/(2f_{\chi}^2)+\cdots\), which will generate multimeson interactions.
The complexity of replacing the meson fields with fermion operators increases with the order of the expansion, as there are no closed-form solutions for such EOMs
\be
\label{eq:mulmeex}
\left(\partial_\mu \partial^\mu+m_\sigma^2\right) \sigma & = &{} - g_{\sigma NN}^{\rm OBE} \bar{N} N \\
& &{} +\sum_{i}g_{\sigma^{i+1}}\sigma^i+\sum_ig_{\sigma^{i+1}NN}\sigma^i\bar{N} N , \nonumber
\ee
where the last two terms involve multimeson interactions. 
Theoretically, such EOMs can be solved iteratively by inputting the Green function solutions, e.g., Eq.~\eqref{eq:OGS}, but they cannot be analytically solved. It seems challenging. But the CSDC rules to be settled on in the following provide a systematic approach.

Since we are working at low energy regions, and if the density does not go very high, the baryonic bilinear currents, like that in Eq.~\eqref{eq:OGS}, should have an order of \( n \propto k_{\rm F}^3\), which can be considered as the characteristic momentum of the density system. For the magnitude of \(k_{\rm F}\) in NM relevant to the cores of massive neutron stars we are interested in, the characteristic momentum is up to \(k_{\rm{c}} \sim 700~{\rm MeV} \sim 10~n_0\), which is the same order as chiral power counting rules. Combining these two kinds of power counting rules, the solution of Eq.~\eqref{eq:mulmeex} can be expanded in a series of \(k_{\rm c}\), which is \(\sim k_{\rm F}\) of Fermi surface and \(\sim p\) of chiral momentum.
In this sense, it can be found that the original Green's function solution in Eq.~\eqref{eq:OGS} has an order of \(k_{\rm{c}}^2\), which can be used as the basis to solve the Eq.~\eqref{eq:mulmeex} iteratively. Note that, here, the couplings are assumed to have a consistent order as the Lagrangians, e.g., the couplings between meson and nucleon is of  \(\mathcal{O}(p) \sim \mathcal{O}(k_c)\) except $g_{\pi NN}^{\rm{OBE}}$ due to its derivative coupling. Then, one can see that \(g_{\sigma^2NN} \sigma \bar{N} N\) is of order \(k_{\rm{c}}^6\), \(g_{\sigma^3NN} \sigma^2 \bar{N} N\) and \(g_{\sigma^3} \sigma^3\) are of order \(k_{\rm{c}}^8\), and so on, so that the solution \(\sigma\sim\mathcal{O}(k_{\rm{c}}^2)+\mathcal{O}(k_{\rm{c}}^4)+\mathcal{O}(k_{\rm{c}}^6)+\cdots\).
If the power of expansion by \(k_{\rm{c}}\) holds, the solution \(\sigma\) can be truncated to the leading order and treated as a induced background at \(\mathcal{O}(k_{\rm{c}}^2)\).
Then, the following assignment can be made: $\sigma,\ \pi,\ \omega_{\mu},\ \rho_{\mu}^i\ \sim \mathcal{O}(k_{\rm{c}}^2)$ for its derivative coupling with the baryons, resulting in the soft pion theorem~\cite{Weinberg:1968de}.

Based on the above discussion of the CSDC rule, one can assign the following order to the Hamiltonian operator of $bs$HLS
\begin{widetext}
\be
\label{eq:HO}
\mathcal{H}_0 & = &	\underbrace{\bar{N}\left(-i \mathbf{\gamma}\cdot\mathbf{\nabla}+m_N\right) N}_{\mathcal{O}(k_{\rm{c}}^4)}\ , \nonumber\\
\mathcal{H}_I & = & \underbrace{g_{\sigma NN}^{\rm OBE} \bar{N} \sigma N+g_{\omega NN}^{\rm OBE} \bar{N} \gamma_\mu \omega^\mu N+g_{\sigma NN}^{\rm OBE} \bar{N} \slashed{\rho}^i \tau^i N+g_{\pi NN}^{\rm OBE} \bar{N} \gamma_5 \slashed{\partial} \pi^i\tau^i N}_{\mathcal{O}(k_{\rm{c}}^6)}-\underbrace{\text{Free Lagrangian of}\ \rho,\ \omega,\ \sigma\ \text{mesons}}_{\mathcal{O}(k_{\rm{c}}^6)} \nonumber\\
& &{} +\underbrace{g_{\sigma^3} \sigma^3+g_{\sigma\rho^2} \sigma \rho^i_\mu \rho^{i,\mu}+g_{\sigma\omega^2} \sigma \omega_{\mu} \omega^{\mu} +g_{\sigma\pi^2} \sigma\pi^i\pi^i+g_{\rho\pi^2} \rho^{\mu,a}\partial_\mu\pi^{b}\pi^{c}\epsilon_{abc}
}_{\mathcal{O}(k_{\rm{c}}^8)} \nonumber\\
& &{} +\underbrace{g_{\sigma^2 N N} \bar{N} \sigma^2 N+g_{\sigma \omega N N} \bar{N} \sigma \omega^\mu \gamma_\mu N+g_{\sigma \rho N N} \bar{N} \sigma \rho^{\mu a} \tau^a \gamma_\mu N}_{\mathcal{O}(k_{\rm{c}}^8)}+ \text{higher order terms}\ .
\ee
\end{widetext}

The relations between the coefficients in Eq.~\eqref{eq:HO} and the parameters in the $bs$HLS model are given after the expansions of \(\chi\) and \(\xi\) fields. The OBE couplings are given by 
\be
  \label{eq:OBEcoup}
		g_{\sigma N N}^{\rm OBE} & = & \frac{m_N}{f_{\chi}}, \, g_{\omega N N}^{\rm OBE} =g_{\omega N N}+g_{\omega N N}^{S S B}, \nonumber\\
        g_{\rho N N}^{\rm OBE} & = & g_{\rho N N}+g_{\rho N N}^{S S B} ,\,  g_{\pi N N}^{\rm OBE}  =g_{\pi N N}+g_{\pi N N}^{S S B} ,
\ee
The other couplings involving more than one mesons are given by
\be
\label{eq:othcoup}
& & g_{\sigma^2 N N} = \frac{m_N}{2 f_\chi^2} , \quad g_{\sigma \omega N N} =g_{\omega N N}^{S S B} \frac{\beta^{\prime}}{f_\chi} ,\nonumber\\
& & g_{\sigma \rho N N} = g_{\rho N N}^{S S B} \frac{\beta^{\prime}}{f_\chi} ,\nonumber\\
& & g_{\sigma^3} = {} -\frac{32 h_5}{3 f_\chi^3}-\frac{\left(4+\beta^{\prime}\right)^3 h_6}{6 f_\chi^3}-m_\pi^2 f_\pi^2 \frac{4}{3 f_\chi^3} ,\nonumber\\
& & g_{\sigma \rho^2} = {} -\frac{m_\rho^2}{f_{\chi}^2} ,\quad g_{\sigma \omega^2} ={} -\frac{m_\omega^2}{f_\chi^2},\nonumber\\
& &\quad g_{\sigma \pi^2} ={} \frac{ m_\pi}{f_\chi^2}, \quad g_{\rho \pi^2} ={} -\frac{ g_\rho}{2 f_\pi^2} .
\ee
The CSDC rule of these couplings can be summarized in Table~\ref{tab:PC}.

\begin{table}[htb]
\centering
\caption{
 The CSDC orders of the mean fields and couplings and parameters in Eqs.~\eqref{eq:LbsHLS}, \eqref{eq:OBEcoup}, \eqref{eq:othcoup}, and \eqref{eq:morecoup} to be given later.
   }
   \label{tab:PC}
   \begin{tabular}{lr|lr|lr}
     \hline
     \hline
     quantity & CSDC &quantity & CSDC & quantity & CSDC  \\
     \hline     
     $m_N$         & $\mathcal{O}(k_c^0)$ 
     &$f_{\pi}$      & $\mathcal{O}(k_c^0)$ 
     &$f_{\chi}$         & $\mathcal{O}(k_c^0)$ \\
     $\Phi$         & $\mathcal{O}(k_c^0)$ 
     &$\beta^\prime$      & $\mathcal{O}(k_c^0)$ 
     &$g^{\rm OBE}_{\pi NN}$         & $\mathcal{O}(k_c^0)$ \\
     $g_A$         & $\mathcal{O}(k_c^0)$  &$g_{\pi NN}$      & $\mathcal{O}(k_c^0)$ &  $g^{\rm SSB}_{\pi NN}$       & $\mathcal{O}(k_c^0)$ \\
          \hline
    $m_{\sigma}$ & $\mathcal{O}(k_c^1)$&$m_{\omega}$& $\mathcal{O}(k_c^1)$& $m_{\rho}$& $\mathcal{O}(k_c^1)$\\
    $m_{\pi}$& $\mathcal{O}(k_c^1)$& $\Phi-1$& $\mathcal{O}(k_c^1)$& $g_\rho$& $\mathcal{O}(k_c^1)$\\
    $g_\omega$& $\mathcal{O}(k_c^1)$& $g^{\rm OBE}_{\sigma NN}$& $\mathcal{O}(k_c^1)$& $g_{\omega NN}$ & $\mathcal{O}(k_c^1)$ \\
    $g^{\rm SSB}_{\omega NN}$ & $\mathcal{O}(k_c^1)$ & $g_{\rho NN}$ & $\mathcal{O}(k_c^1)$ &
    $g^{\rm OBE}_{\omega NN}$& $\mathcal{O}(k_c^1)$ \\
    $g^{\rm SSB}_{\rho NN}$ & $\mathcal{O}(k_c^1)$& $g^{\rm OBE}_{\rho NN}$& $\mathcal{O}(k_c^1)$ & $g_{\sigma^2 NN}$& $\mathcal{O}(k_c^1)$ \\
    $g_{\sigma\omega NN}$& $\mathcal{O}(k_c^1)$& $g_{\sigma\rho NN}$& $\mathcal{O}(k_c^1)$ & $g_{\sigma\pi^2}$& $\mathcal{O}(k_c^1)$ \\
    $g_{\sigma^3 NN}$ & $\mathcal{O}(k_c^1)$& $g_{\rho\pi^2}$& $\mathcal{O}(k_c^1)$ & $g_{\sigma^2\omega NN}$& $\mathcal{O}(k_c^1)$ \\ $g_{\sigma^2\rho NN}$& $\mathcal{O}(k_c^1)$& $g_{\sigma^4 N N}$ & $\mathcal{O}(k_c^1)$ & $g_{\sigma^3\omega NN}$& $\mathcal{O}(k_c^1)$ \\
    $g_{\sigma^3\rho NN}$& $\mathcal{O}(k_c^1)$& $g_{\sigma^3 NN}$ & $\mathcal{O}(k_c^1)$   \\
         \hline
    $\sigma$& $\mathcal{O}(k_c^2)$& $\omega$& $\mathcal{O}(k_c^2)$& $\rho$& $\mathcal{O}(k_c^2)$\\
    $\pi$& $\mathcal{O}(k_c^2)$& $h_5$& $\mathcal{O}(k_c^2)$& $h_6$& $\mathcal{O}(k_c^2)$\\
    $g_{\sigma\rho^2}$& $\mathcal{O}(k_c^2)$& $g_{\sigma^3}$& $\mathcal{O}(k_c^2)$&  $g_{\sigma\omega^2}$& $\mathcal{O}(k_c^2)$\\
    $g_{\sigma^4}$& $\mathcal{O}(k_c^2)$ &     $g_{\sigma^2\omega^2}$
    & $\mathcal{O}(k_c^2)$&$g_{\sigma^2\rho^2}$& $\mathcal{O}(k_c^2)$\\
    $g_{\sigma^5}$& $\mathcal{O}(k_c^2)$&    $g_{\sigma^3\omega^2}$& $\mathcal{O}(k_c^2)$&    $g_{\sigma^3\rho^2}$ & $\mathcal{O}(k_c^2)$\\
     \hline
     \hline
   \end{tabular}
 \end{table}

\begin{figure}[htbp]
	\centering
	\includegraphics[width=0.45\textwidth]{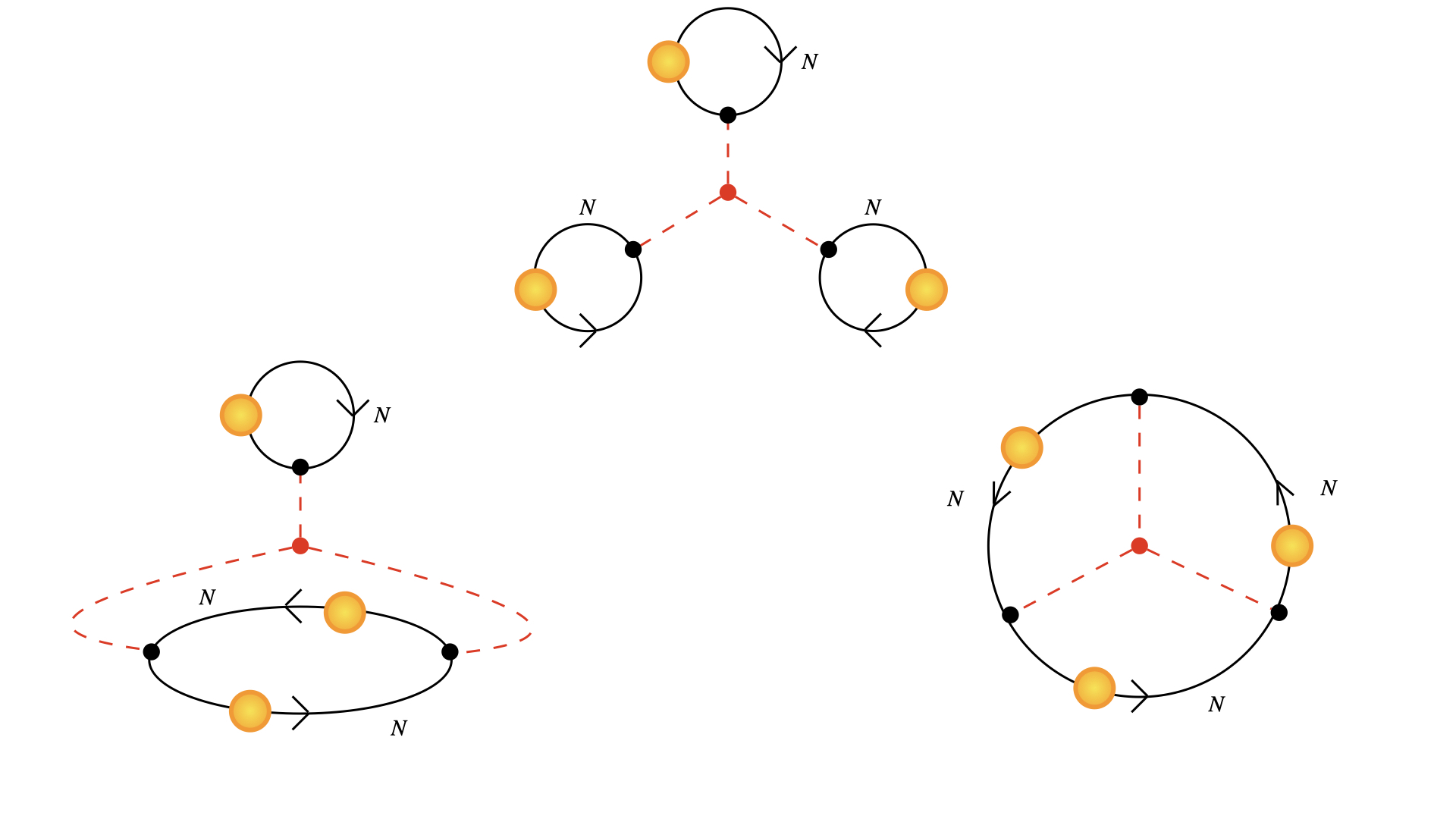}
	\caption{
    The effective Feynman diagrams for three-meson terms in Hamiltonian.
    The red points indicate the possible three-meson interaction terms, e.g., $\sigma^3$, $\sigma\omega\omega $, $\sigma\rho\rho$, and $\sigma\pi\pi$.
  }
	\label{fig:3meson}
\end{figure}

As a result, the computation of the operators containing multimeson fields is equivalent to the calculation of the fermion loops in the Feynman diagrams. An example from three-meson terms are shown in Fig.~\ref{fig:3meson}.
It is worth noting that the LO, [$\mathcal{O}(k_{\rm{c}}^4)$], contribution to the EOS is from free Fermi fluid, the NLO,  [$\mathcal{O}(k_{\rm{c}}^6)$], is the OBE-type interaction and those with more than one mesons are of higher order,  [$\geq \mathcal{O}(k_{\rm{c}}^8)$].

\subsection{The CSDC at finite temperature and nuclear matter properties under RMF approximation}
\label{subsec:TCSDC}

Previously, the CSDC rules were established at zero temperature within RHF method.
In order to study the thermal properties of NM beyond OBE, \(\mathcal{O}(k_{\rm{c}}^6)\) in the CSDC rules, the computation of multiloop diagrams, such as in Fig.~\ref{fig:3meson}, will be a great challenge.
To avoid this complexity and see the effectiveness of the CSDC rules in studies on thermal properties of NM, we resort to the RMF approximation~\cite{Walecka:1974qa}, where exchange (Fock) terms are neglected, e.g., only three-tadpole diagram in Fig.~\ref{fig:3meson} is kept.
In the RMF approximation, the meson fields are treated as ``classical'' background fields induced by the baryon sources and the matter is assumed to be homogeneous.

In addition, during the extension to finite temperature, it is convenient to introduce the Landau potential~\cite{Kapusta:2006pm}, \(\Phi_{\rm G}\), compared to Hamiltonian discussed in previous sections. Here, we again use the OBE model in Eq.~\eqref{eq:LOBE} as an example to illustrate the procedure. Under RMF approximation, the Lagrangian in Eq.~\eqref{eq:LOBE} can be simplified as
\be
	\mathcal{L}_{\rm RMF} & = & \bar{N}\left[i \slashed \partial-m_N -g_{\omega N N}^{\mathrm{OBE}} \gamma_0\omega \right. \nonumber\\
    & & \left. {}\quad -g_{\rho N N}^{\mathrm{OBE}} \gamma_0\rho\tau_3-g_{\sigma N N}^{\mathrm{OBE}} \sigma\right]N + \mathcal{L}_{\rm M} ,
\ee
where only the time components of vector meson omega and charge-neutral rho meson, and scalar meson \(\sigma\) survive.
\(\mathcal{L}_{\rm M}\) is the RMF simplification of \(\mathcal{L}_{\mathrm{meson}}\). Then, \(\Phi_{\rm G}\) can be obtained via generating functional as~\cite{Kapusta:2006pm},
\begin{equation}
	\Phi_{\mathrm{G}}={}-T \ln Z .
\end{equation}
Here, \(Z\) is the grand partition function of the system which has the path-integral representation
\begin{widetext}
\be
Z & = & \int\left[\mathrm{d} \bar{N}_p\right]\left[\mathrm{d} N_p\right]\left[\mathrm{d} \bar{N}_n\right]\left[\mathrm{d} N_n\right] \exp \left(\int_0^\beta \mathrm{d} \tau \int \mathrm{d}^{3}x\left(\mathcal{L}_{\mathrm{RMF}}+\mu_p N_p^{\dagger} N_p+\mu_n N_n^{\dagger} N_n\right)\right) ,
\ee
where \(\mu_p\) and \(\mu_n\) are the chemical potentials of protons and neutrons, respectively, \(\beta=1/T\) is the inverse temperature, and \(N_{p(n)}\) refers to the proton (neutron) field. The integral is over state space of protons and neutrons, and mesons are treated as classical background fields in the RMF approximation. \(N_{p(n)}^\dagger N_{p(n)}\) is the proton (neutron) number density operator.
With the homogeneous condition, the partition function finally arrives at
\be
\ln Z & = & \sum_{i=n, p} \frac{V}{T} \int \frac{\mathrm{d}\bold{p}}{(2 \pi)^3}\left\{2 E^*_i(\bold{p})+2 T \ln \left(1+\exp \left(-\frac{E^*_i(\bold{p})-\mu^*_i}{T}\right)\right) + 2 T \ln \left(1+\exp \left(-\frac{E^*_i(\bold{p})+\mu^*_i}{T}\right)\right)\right\} \nonumber\\
& &{} +\frac{V}{T} \mathcal{L}_{\mathrm{M}} ,
\ee
\end{widetext}
where \(E^*_i(\bold{p})=\sqrt{\bold{p}^2+m_i^{*2}}\) is the effective energy of nucleons with effective mass \(m_{p(n)}^*=m_N+g_{\sigma N N}^{\mathrm{OBE}} \sigma\), $i$ is the flavor index, and effective chemical potential {\(\mu_{p(n)}^*=\mu_{p(n)}-g_{\omega N N}^{\mathrm{OBE}} \omega\mp g_{\rho N N}^{\mathrm{OBE}}\rho\).}
The first term in the integral kernel will be removed by assuming the vacuum energy is zero, the same effect as normal ordering in Eq.~\eqref{eq:HG}, after the normalization process.
The EOS of homogeneous NM can then be calculated via Landau potential density, \(\Omega=\Phi_{\rm G}/V\), and 
\be
P & = &{} -\Omega ,\ \nonumber\\
n_i & = & \left.-\frac{\partial \Omega}{\partial \mu_i}\right|_{T} ,\nonumber\\
s & = & \left.-\frac{\partial \Omega}{\partial T}\right|_{\mu_p,\mu_n} ,\nonumber\\ \epsilon & = & \Omega+ \sum_i \mu_i n_i+Ts ,
\ee
where \(P\), \(n_i\), \(s\), and \(\epsilon\) are the pressure, particle number density, entropy density and energy density of the system, respectively.
In addition, the RMF EOMs of the meson fields can be obtained by functional variation of \(Z\), or equivalently \(\Omega\), with respect to the meson fields, e.g., \(\partial \Omega/\partial \sigma=0\) for the sigma field.

When \(T\) approaches \(0\) ($\beta\to\infty$), the baryon densities
\be
\label{eq:rhoT}
n_i & = & 2\int \frac{{\rm d}\bold{p}}{(2\pi)^{3}}\!\left[\frac{1}{e^{\beta(E^{*}_i-\mu^{*}_i)}+1}-\frac{1}{e^{\beta(E^{*}_i+\mu^{*}_i)}+1}\right] \nonumber\\
& \to & 2\int \frac{{\rm d}\bold{p}}{(2\pi)^{3}}\!\left[\theta(\mu^{*}_i-E^{*}_i)-\theta(-\mu^{*}_i-E^{*}_i)\right] ,
\ee
where the former and latter terms in the integral kernel account for the occupied and unoccupied states, respectively.
We can see that \(\mu_i^*=0\) will lead to \(n_i=0\), and assume the background meson fields also vanish consistently, as the physical vacuum condition.
If the interactions between nucleons can be treated as perturbations, as in low-energy EFTs of QCD, the contributions of meson fields to the \(m_{p(n)}^*\) and \(\mu_{p(n)}^*\) can be also regarded as perturbations when \(\mu_{p(n)}^*\) is not far away from vacuum value, so that \(\mu_i^*\sim \mu_i\) and \(m_i^*\sim m_N\).
Then, only occupied states contribute to the integral in Eq.~\eqref{eq:rhoT} for NM at \(T=0\), due to the fact that \(\mu_{p(n)}^*\sim \mu_{p(n)}\) and \(E_{p(n)}^*\gtrsim m_N \) are both positive with above considerations, which makes the no-sea approximation in Eq.~\eqref{eq:psiB} reasonable for the description of zero-temperature NM. And \(\mu_i\in[0,m_N]\) will lead to \(n_i=0\).
So, the familiar density expression we used at zero temperature in Eq.~\eqref{eq:psiB} can be obtained
\begin{equation}
	n_i=\frac{1}{3\pi^{2}}\,k_{{\rm F}, i}^{3} ,
\end{equation}
where the Fermi momentum \(k_{{\rm F}, i}\) is defined by \(\mu^{*}_i=\sqrt{\bold{k}_{{\rm F}, i}^{2}+m_{i}^{*2}}\).
This not only explains the reason of expanding the nucleon fields within \(\bold{k}_{{\rm F}}\), but also shows that the extension of the CSDC rules to finite temperature is possible with cares about the temperature effect on the Fermi surface and the occupation of single-particle states.

We notice that the integral kernel in Eq.~\eqref{eq:rhoT} changes much, when \(\bold{p}\) makes \(|E^*_i-\mu^{*}_i|<T\), if \(T\) is small enough. In this case, the contribution of unoccupied states starts but is still significantly suppressed by the factor, \(|E^*_i+\mu^{*}_i|\) with above perturbation assumptions.

\begin{figure}[htbp]
\centering
\includegraphics[width=0.45\textwidth]{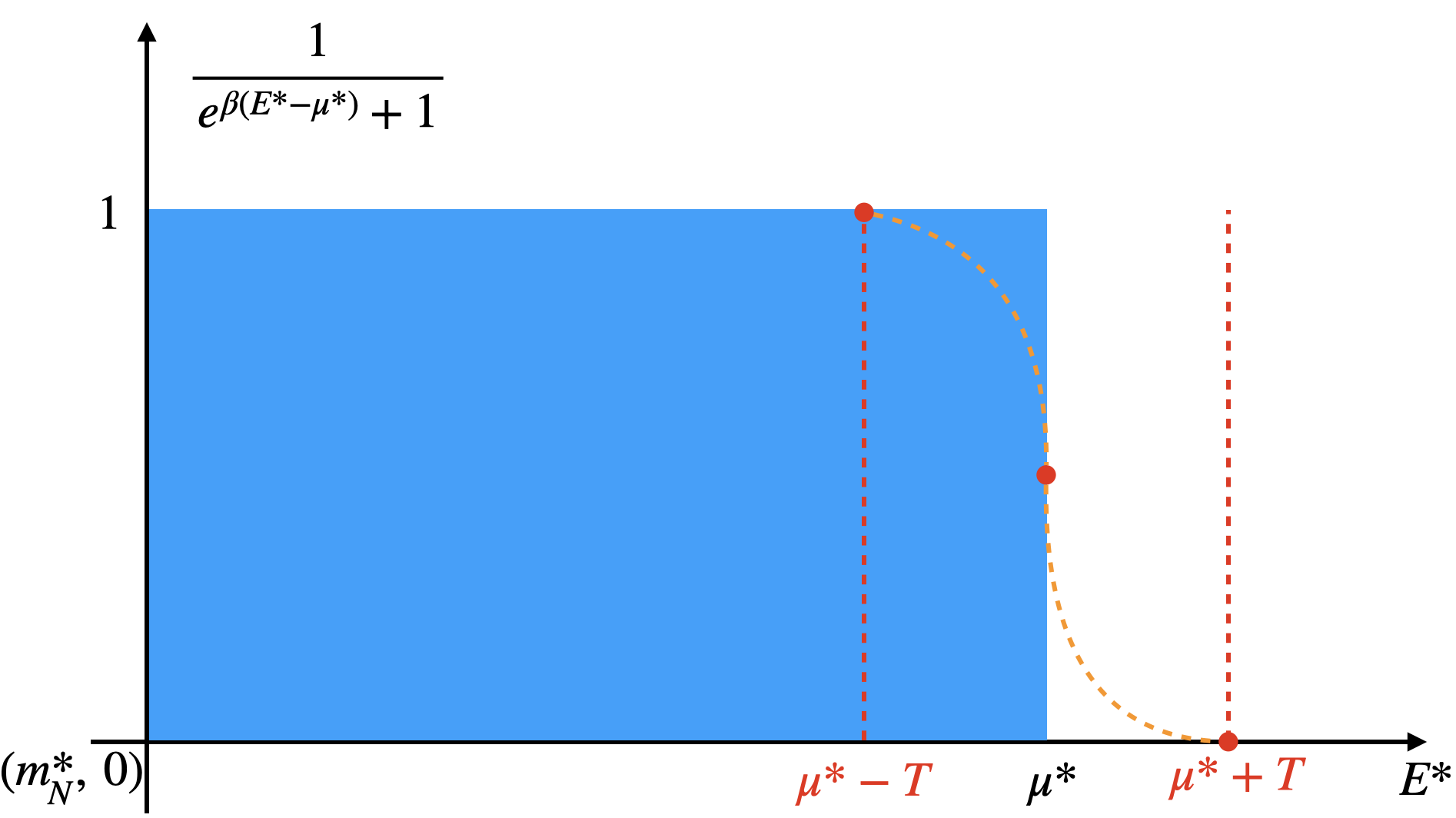}
\includegraphics[width=0.45\textwidth]{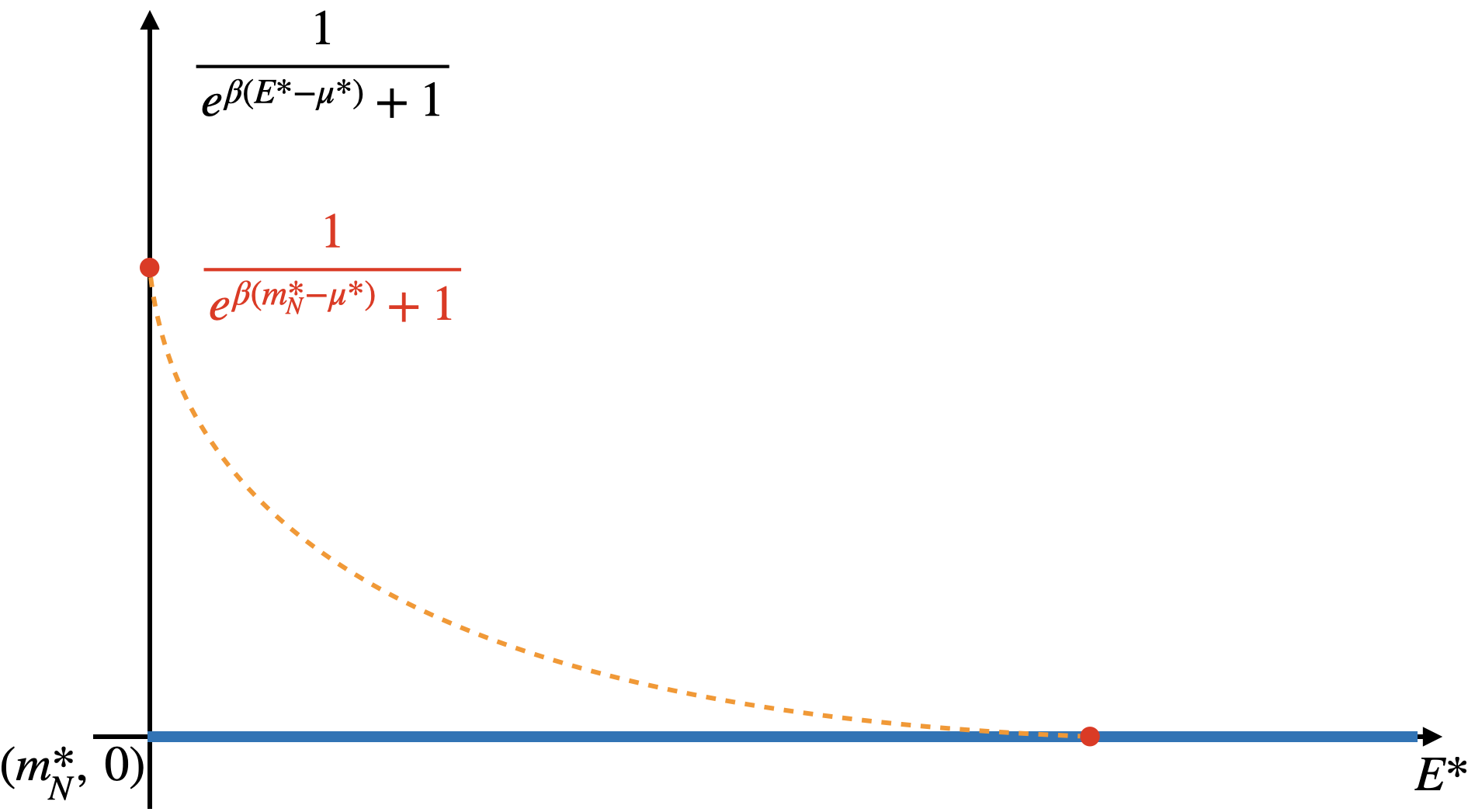}
\caption{
    The most contributed integral kernel in Eq.~\eqref{eq:rhoT} with perturbation assumptions for \(\mu_i^*>m_N^*\) (upper panel) and \(\mu_i^*<m_N^*\) (lower panel).
    The blue shaded region indicates the kernel at \(T=0\).
    The orange dashed lines is the shift of the integral kernel at finite \(T\).
    }
\label{fig:rhoD}
\end{figure}

As shown in Fig.~\ref{fig:rhoD}, CSDC rules applicable region should be modified by the estimation
\begin{enumerate}
    \item When \(\mu_i^*>m_i^*\)
    \be
  \label{eq:largemuD}
  \tilde{k}_{\mathrm{F}} & = & \sqrt{(\mu^{*}_i+T)^2-m_{i}^{*2}} \nonumber\\
  & \approx & \sqrt{(\mu_i+T)^2-m_N^2}\lesssim m_N , \mbox{~for~~}  T\ll \mu_i^*\approx\mu_i ,
  \nonumber\\
\ee
due to the fact the integral result of Eq.~\eqref{eq:rhoT} will be smaller than \(\frac{2}{3\pi^{2}}\,\tilde{k}_{\mathrm{F}}^{3}\), which should stay at the same order of chiral order;

\item  When \(\mu_i^*<m_i^*\)

\begin{equation}
  \label{eq:smallmuD}
	 \frac{1}{e^{\beta(m_i^*-\mu_i^*)}+1}\approx \frac{1}{e^{\beta(m_N-\mu_i)}+1}\ll 1 ;
\end{equation}
\end{enumerate}
Above conditions give regions in the \(T-\mu_i\) plane where CSDC rules are applicable.

In practice, the effects of temperature on the integral kernel make it deviate to the \(T=0\) case. 
For \(\mu_i>m_N\), the deviation can be estimated by \(T/\mu_i\lesssim \delta\), while for \(\mu_i<m_N\), the deviation can be estimated by 
\(1/(e^{\beta(m_i^*-\mu_i^*)}+1)\lesssim \delta\). 
Then, it should be noted that \(\mu_i=\partial\epsilon/\partial n_i|_{s}\sim \mathcal{O}{(k_{\rm{c}})}+\mathcal{O}{(k_{\rm{c}}^3)}+...\), as a result \(\mu_i^* \sim \mathcal{O}{(k_{\rm{c}})}+\mathcal{O}{(k_{\rm{c}}^3)}+...\), where orders higher than one are corrections from interactions.
The same reasoning applies to the effective mass and Landau potential density at low temperature.  One more thing should be noted is that \(\mu_i\in[0,m_N]\) corresponds to \(n_i=0\) at \(T=0\), but the \(n_i\approx 0\) region will shrink to \(\mu_i\in[0,m_N-T]\) with \(T\) increasing, with the same reasoning in Eq.~\eqref{eq:smallmuD}.

\section{Nuclear matter properties with CSDC}
\label{sec:res}

After establishing the CSDC rules for NM at finite density and temperature, we now turn to the NM matter properties at densities and temperatures. The thermal quantities of NM are calculated up to \(\mathcal{O}(k_{\rm{c}}^{12})\) in the CSDC rules.

In the numerical calculation, we take the empirical values \(m_N=939~\mathrm{MeV}\), \(m_{\pi}=135~\mathrm{MeV}\), \(m_\omega=782~\mathrm{MeV}\), and \(m_\rho=763~\mathrm{MeV}\) according to Particle Data Group~\cite{ParticleDataGroup:2024cfk}, \(f_{\pi}=91.9~\mathrm{MeV}\) from Ref.~\cite{Aoki:2016frl} as inputs. Then, \(h_{5}\) and $h_6$ are fixed by Eq.~\eqref{eq:sp}. The relevant couplings \(g_{\omega NN}^{\rm OBE}, g_{\rho NN}^{\rm OBE}\), \( g_{\sigma\omega NN}\) and \(g_{\sigma\rho NN}\) defined in Eq.~\eqref{eq:OBEcoup}, and \(f_{\chi}\), \(\beta'\), and \(m_\sigma\) are pinned down using the NM properties around saturation density list in Table~\ref{tab:RES}.
During the fitting process, the parameter $f_\chi$ is taken around $f_\pi$~\cite{Crewther:2013vea}, with $\beta^{\prime}$ generally considered to lie within the range $1 \sim 2$ ~\cite{Shao:2022njr}.
Based on these considerations, we set the parameter $f_\chi$ to $80 \sim 300 \mathrm{MeV}$, while $\beta^{\prime}$ is varied within $0.5 \sim 3$. \(m_{\sigma}\) is considered to be $600 \pm 200~ \mathrm{MeV}$~\cite{ParticleDataGroup:2024cfk}, to consistent with the value at vacuum.

In addition to those defined in Eq.~\eqref{eq:othcoup}, the other couplings involved in the calculation can be estimated from the above free parameters through relations
\be
\label{eq:morecoup}
g_{\sigma^3 NN} & = & \frac{m_N}{6 f_{\chi}^{3}} , \quad g_{\sigma^4 NN} = \frac{m_N}{24 f_{\chi}^{4}} , \nonumber\\
g_{\sigma^3\omega NN} & = & \frac{\beta'^{2} g_{\sigma\omega NN}}{6 f_{\chi}^{2}} , \quad g_{\sigma^3\rho NN} = \frac{\beta'^{2} g_{\sigma\rho NN}}{6 f_{\chi}^{2}}, \nonumber\\
g_{\sigma^2\omega^2} & = &{} -\frac{m_\omega^{2}}{f_{\chi}^{2}}\ , \quad g_{\sigma^2\rho^2} ={} -\frac{m_\rho^{2}}{f_{\chi}^{2}} ,\nonumber\\
g_{\sigma^3\omega^2} & = &{} -\frac{2 m_\omega^{2}}{3 f_{\chi}^{3}} , \quad g_{\sigma^3\rho^2} ={} -\frac{2 m_\rho^{2}}{3 f_{\chi}^{3}}, \nonumber\\
g_{\sigma^4} & = &{} -\frac{4^{4} h_5 + (4+\beta')^{4} h_6 + 2^{4} m_{\pi}^2f_{\pi}^2}{24 f_{\chi}^{4}} , \nonumber\\
g_{\sigma^5} & = &{} -\frac{4^{5} h_5 + (4+\beta')^{5} h_6 + 2^{5} m_{\pi}^2f_{\pi}^2}{120 f_{\chi}^{5}} .
\ee
Our results of the parameters are summarized in Table~\ref{tab:CCP}. As a comparison, we also list the values of the parameters used in TM1~\cite{Sugahara:1993wz}.

\subsection{EOS and bulk properties around saturation density}

At first, let us consider the symmetric nuclear matter (SNM) around saturation density. It should be noted that, after CSDC, the effective mass and chemical potentials have different expressions order by order. Explicitly
\begin{itemize}
\item At order $\mathcal{O}(k_{\mathrm{c}}^{6})$, 
\begin{subequations}
\be
m_N^* & = & m_N + g^{\rm OBE}_{\sigma NN} \sigma , \\
\mu_p^* & = & \mu_p - g^{\rm OBE}_{\omega NN} \omega - g^{\rm OBE}_{\rho NN} \rho , \\
\mu_n^* &= & \mu_n - g^{\rm OBE}_{\omega NN} \omega + g^{\rm OBE}_{\rho NN} \rho ;
\ee
\end{subequations}

\item At order $\mathcal{O}(k_{\mathrm{c}}^{8})$,
\begin{subequations}
\be
m_N^* & = & m_N + g^{\rm OBE}_{\sigma NN} \sigma + g_{\sigma^2 NN} \sigma^2 , \\
\mu_p^* & = & \mu_p - \left(g^{\rm OBE}_{\omega NN} + g_{\sigma\omega NN} \sigma\right) \omega \nonumber\\
& &{} - \left(g^{\rm OBE}_{\rho NN} + g_{\sigma\rho NN} \sigma\right) \rho , \\
\mu_n^* & = & \mu_n - \left(g^{\rm OBE}_{\omega NN} + g_{\sigma\omega NN} \sigma\right) \omega \nonumber\\
& &{} + \left(g^{\rm OBE}_{\rho NN} + g_{\sigma\rho NN} \sigma\right) \rho ;
\ee
\end{subequations}

\item At order $\mathcal{O}(k_{\mathrm{c}}^{10})$,
\begin{subequations}
\be
m_N^* & = & m_N + g^{\rm OBE}_{\sigma NN} \sigma + g_{\sigma^2 NN} \sigma^2 + g_{\sigma^3 NN} \sigma^3, \\
\mu_p^* & = & \mu_p - \left(g^{\rm OBE}_{\omega NN} + g_{\sigma\omega NN} \sigma + g_{\sigma^2\omega NN} \sigma^2\right) \omega \nonumber\\
& &{} - \left(g^{\rm OBE}_{\rho NN} + g_{\sigma\rho NN} \sigma + g_{\sigma^2\rho NN} \sigma^2\right) \rho , \\
\mu_n^* & = & \mu_n - \left(g^{\rm OBE}_{\omega NN} + g_{\sigma\omega NN} \sigma + g_{\sigma^2\omega NN} \sigma^2\right) \omega \nonumber\\
& &{} + \left(g^{\rm OBE}_{\rho NN} + g_{\sigma\rho NN} \sigma + g_{\sigma^2\rho NN} \sigma^2\right) \rho ;
\ee
\end{subequations}

\item At order $\mathcal{O}(k_{\mathrm{c}}^{12})$,
\begin{subequations}
\be
m_N^* & = & m_N + g^{\rm OBE}_{\sigma NN} \sigma \nonumber\\
& &{} + g_{\sigma^2 NN} \sigma^2 + g_{\sigma^3 NN} \sigma^3 + g_{\sigma^4 NN} \sigma^4, \\
\mu_p^* & = & \mu_p - \left(g^{\rm OBE}_{\omega NN} + g_{\sigma\omega NN} \sigma \right. \nonumber\\
& &\left.{} \qquad \quad + g_{\sigma^2\omega NN} \sigma^2 + g_{\sigma^3\omega NN} \sigma^3\right) \omega\nonumber \\
& &{} \quad - \left(g^{\rm OBE}_{\rho NN} + g_{\sigma\rho NN} \sigma \right. \nonumber\\
& &\left.{} \qquad \quad + g_{\sigma^2\rho NN} \sigma^2 + g_{\sigma^3\rho NN} \sigma^3\right) \rho , \\
\mu_n^* & = & \mu_n - \left(g^{\rm OBE}_{\omega NN} + g_{\sigma\omega NN} \sigma \right. \nonumber\\
& &\left.{} \qquad \quad + g_{\sigma^2\omega NN} \sigma^2 + g_{\sigma^3\omega NN} \sigma^3\right) \omega\nonumber \\
& &{} \quad + \left(g^{\rm OBE}_{\rho NN} + g_{\sigma\rho NN} \sigma \right. \nonumber\\
& &\left.{} \qquad \quad + g_{\sigma^2\rho NN} \sigma^2 + g_{\sigma^3\rho NN} \sigma^3\right) \rho .
\ee
\end{subequations}
\end{itemize}
For SNM, \(\mu_n=\mu_p\) and \(\mu_p^*=\mu_n^*\) due to exact isospin symmetry.

\begin{table}[htb]
  \caption{
    Nuclear matter properties around saturation density $n_0$. $E(n_0)$, $E_{\text {sym }}(n_0)$, $L(n_0)$, $K(n_0)$ are binding energy, symmetry energy, symmetry energy slope and incompressibility ~\cite{Danielewicz:2002pu} at $n_0$. Their definitions can be found in Ref.~\cite{Ma:2025llw}.
    \(T_{\rm c}\) is the GLPT critical temperature.
    $E(n_a)$ is binding energy at $n_a=1.5 n_0$. $n_0$ is in unit of fm$^{-3}$ and others are in unit of MeV.
  }
  \label{tab:RES}
  \begin{tabular}{lccccc}
    \hline
    \hline
    &Empirical & $\mathcal{O}(k_{\mathrm{c}}^{6})$ & $\mathcal{O}(k_{\mathrm{c}}^{8})$ & $\mathcal{O}(k_{\mathrm{c}}^{10})$ & $\mathcal{O}(k_{\mathrm{c}}^{12})$ \\
    \hline
      $n_0$ & $0.155 \pm 0.050$~\cite{Ma:2025llw} & $0.160$ & $0.160$ & $0.160$ & $0.158$ \\
      $E(n_0)$ & $-15.0 \pm 1.0 $~\cite{Sedrakian:2022ata} & $-16.3$ & $ -15.6$ & $-16.1$ & $-15.4$ \\
      $E(n_a)$ & $-13.3 \pm 0.5 $ ~\cite{Leifels:2015iei} & $-6.92$ & $-12.0$ & $-8.99$ & $-11.2$ \\
      $K(n_0)$ & $230 \pm 30$ ~\cite{Dutra:2012mb} & $566$ & $366$ & $572$ & $391$ \\
      $E_{\mathrm{sym}}(n_0)$ & $30.9 \pm 1.9 $~\cite{lattimer2013constraining} & $31.6$ & $32.1$ & $ 31.8$ & $32.2$ \\
      $L(n_0)$ & $52.5 \pm 17.5$~\cite{lattimer2013constraining} & $104$ & $83.2$ & $ 75.7$ & $77.8$ \\
      $T_{\rm c}$ & \(20.0\pm3.0\)~\cite{Karnaukhov:2003vp} & $19.0$ & $26.5$ & $24.0$ & $22.5$ \\
    \hline
    \hline
  \end{tabular}
\end{table}
\begin{table}[htb]
  \centering
  \caption{
    The values of free parameters and couplings.
    The couplings of relevant couplings at each order and TM1~\cite{Sugahara:1993wz} are also listed. 
  }
  \label{tab:CCP}
  \begin{tabular}{lrrrrr}
    \hline
    \hline
    CSDC order & $\mathcal{O}(k_{\rm{c}}^6)$ & $\mathcal{O}(k_{\rm{c}}^8)$ & $\mathcal{O}(k_{\rm{c}}^{10})$ & $\mathcal{O}(k_{\rm{c}}^{12})$& TM1 \\
    \hline
    $m_{\sigma}\,\left(\mathrm{MeV}\right)$      & $588$ & $421$ & $745$  & $379$& $511$ \\
    $f_{\chi}\,\left(\mathrm{MeV}\right) $       & $-82.2$ & $-264$  & $-122$   & $-277$  & $\cdots$ \\
    $\beta^\prime$                               & $\cdots$ & $1.92$  & $0.744$   & $0.910$&$\cdots$ \\
    $h_5\,\left(10^{9}\,\mathrm{MeV}^{4}\right)$                             & $\cdots$ & $1.47$  & $2.51$   & $3.45$&$\cdots$ \\
    $h_6\,\left(10^{9}\,\mathrm{MeV}^{4}\right)$                             & $\cdots$ & $-1.04$ & $ -2.18$ & $-2.96$ & $\cdots$ \\
    $g^{\rm OBE}_{\sigma NN}$                              & $-11.4$ & $-3.54$  & $-7.68$   & $-3.39$  & $10.0$\\
    $g^{\rm OBE}_{\omega NN}$                              & $13.2$  & $-1.50$  & $4.14$    & $0.311$ &$12.6$ \\
    $g^{\rm OBE}_{\rho NN}$                                & $3.27$   & $-4.24$  & $-4.54$   & $-4.86$ & $4.63$ \\
    $g_{\sigma^2 NN}\,\left(10^{-3}\,\mathrm{MeV}^{-1}\right)$ & $\cdots$      & $6.69$   & $31.4$    & $6.11$& $\cdots$ \\
    $g_{\sigma\omega NN}\,\left(10^{-1}\,\mathrm{MeV}^{-1}\right)$ & $\cdots$      & $2.00$   & $0.681$   & $1.23$& $\cdots$ \\
    $g_{\sigma\rho NN}\,\left(10^{-2}\,\mathrm{MeV}^{-1}\right)$   & $\cdots$      & $1.44$   & $5.46$    & $4.16$& $\cdots$ \\
    $g_{\sigma^3}\,\left(10^{3}\,\mathrm{MeV}\right)$    & $\cdots$ & $-1.09$  & $-6.46$   & $-0.754$ & $-0.473$\\
    $g_{\sigma\rho^2}\,\left(10^{3}\,\mathrm{MeV}\right)$        & $\cdots$      & $2.20$   & $4.76$    & $2.10$ & $0$ \\
    $g_{\sigma\omega^2}\,\left(10^{3}\,\mathrm{MeV}\right)$    & $\cdots$      & $2.31$   & $5.00$    & $2.21$ & $0$ \\
    $g_{\sigma^3 NN}\,\left(10^{-5}\,\mathrm{MeV}^{-2}\right)$ & $\cdots$   & $\cdots$      & $-8.55$   & $-0.734$ & $0$\\
    $g_{\sigma^2\omega NN}\,\left(10^{-5}\,\mathrm{MeV}^{-2}\right)$  & $\cdots$   & $\cdots$      & $-20.7$   & $-20.2$ & $0$\\
    $g_{\sigma^2\rho NN}\,\left(10^{-5}\,\mathrm{MeV}^{-2}\right)$    & $\cdots$   & $\cdots$      & $-16.6$   & $-6.82$ & $0$\\
    $g_{\sigma^4}$                     & $\cdots$      & $\cdots$      & $85.5$   & $4.51$ & $0.152$\\
    $g_{\sigma^2\omega^2}$                     & $\cdots$      & $\cdots$      & $-40.9$  & $-7.96$ & $\cdots$ \\
    $g_{\sigma^2\rho^2}$                         & $\cdots$      & $\cdots$      & $-38.9$  & $-7.57$  & $\cdots$ \\
    $g_{\sigma^4 N N} (10^{-9} \mathrm{MeV}^{-3})$ & $\cdots$ & $\cdots$      & $\cdots$       & $6.62$  & $\cdots$ \\
    $g_{\sigma^3\omega NN}(10^{-7} \mathrm{MeV}^{-3})$    & $\cdots$ & $\cdots$      & $\cdots$  & $2.21$  & $\cdots$ \\
    $g_{\sigma^3\rho NN}(10^{-8} \mathrm{MeV}^{-3})$      & $\cdots$ & $\cdots$  & $\cdots$       & $7.47$  & $\cdots$ \\
    $g_{\sigma^5}(10^{-2} \mathrm{MeV}^{-1})$ & $\cdots$ & $\cdots$  & $\cdots$  & $-1.93$  & $\cdots$ \\
    $g_{\sigma^3\omega^2}(10^{-2} \mathrm{MeV}^{-1})$ & $\cdots$ & $\cdots$      & $\cdots$       & $1.91$  & $\cdots$ \\
    $g_{\sigma^3\rho^2}(10^{-2} \mathrm{MeV}^{-1})$     & $\cdots$ & $\cdots$      & $\cdots$       & $1.82$  & $\cdots$ \\
    $g_{\omega^4}$     & $\cdots$ & $\cdots$      & $\cdots$       & $\cdots$  & $-17.8$ \\
    \hline
    \hline
  \end{tabular}
\end{table}

\begin{figure}[tbh]
    \centering
    \includegraphics[width=0.45\textwidth]{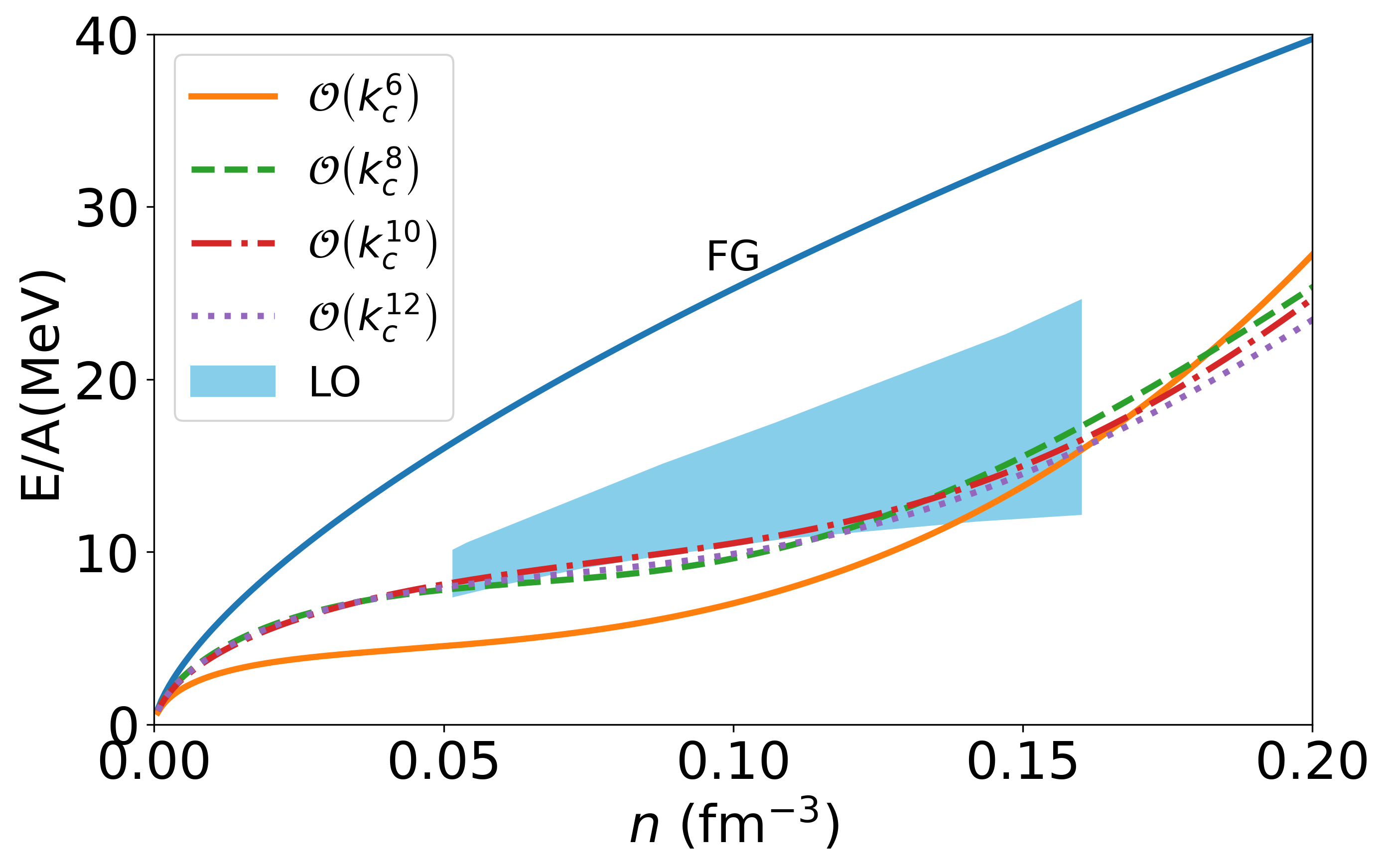}
    \caption{
      Energy per nucleon as a function of nucleon number density for PNM.
      FG refers to free Fermi gas, the leading order in CSDC, \(\mathcal{O}(k_{\rm{c}}^4)\).
      The blue shaded region represents the results from the leading-order chiral nuclear-force~\cite{Drischler:2021kxf}.
    }
    \label{fig:EPN_rho_PNM}
\end{figure}

Our results are summarized in Table~\ref{tab:RES}, with the fitted parameters shown in Table~\ref{tab:CCP}. From Table~\ref{tab:CCP} and Eqs.~\eqref{eq:OBEcoup},~\eqref{eq:othcoup}, and~\eqref{eq:morecoup}, we can see that there are no more free parameters beyond \(\mathcal{O}(k_{\rm{c}}^8)\) due to the chiral-scale expansion. We also compare the EOSs for pure neutron matter (PNM) of our results with that from the leading-order chiral nuclear-force (\(\chi\)NF)~\cite{Drischler:2021kxf} in Fig.~\ref{fig:EPN_rho_PNM}.

\begin{figure}[tbh]
  \centering
    \includegraphics[width=0.45\textwidth]{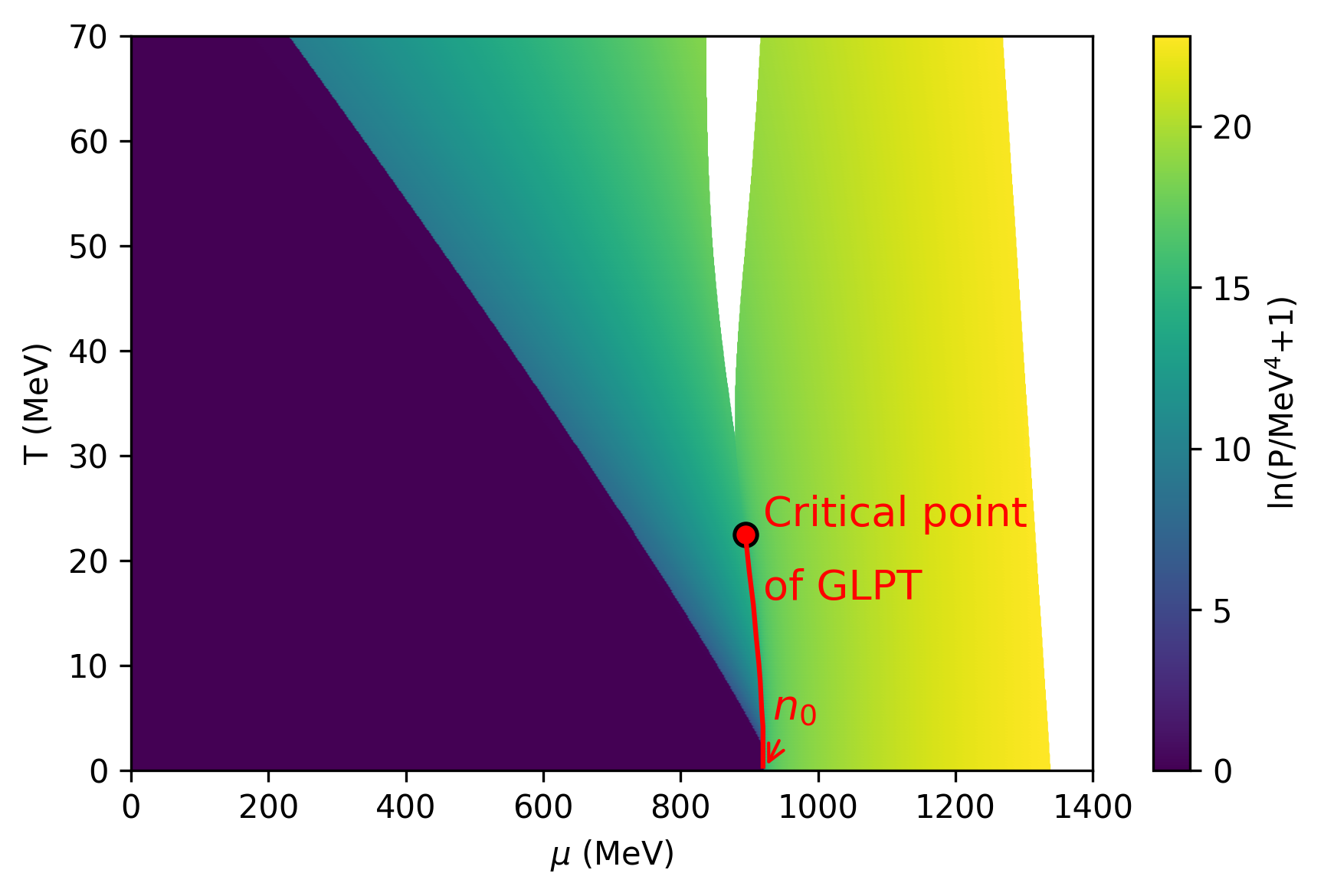}
    \includegraphics[width=0.45\textwidth]{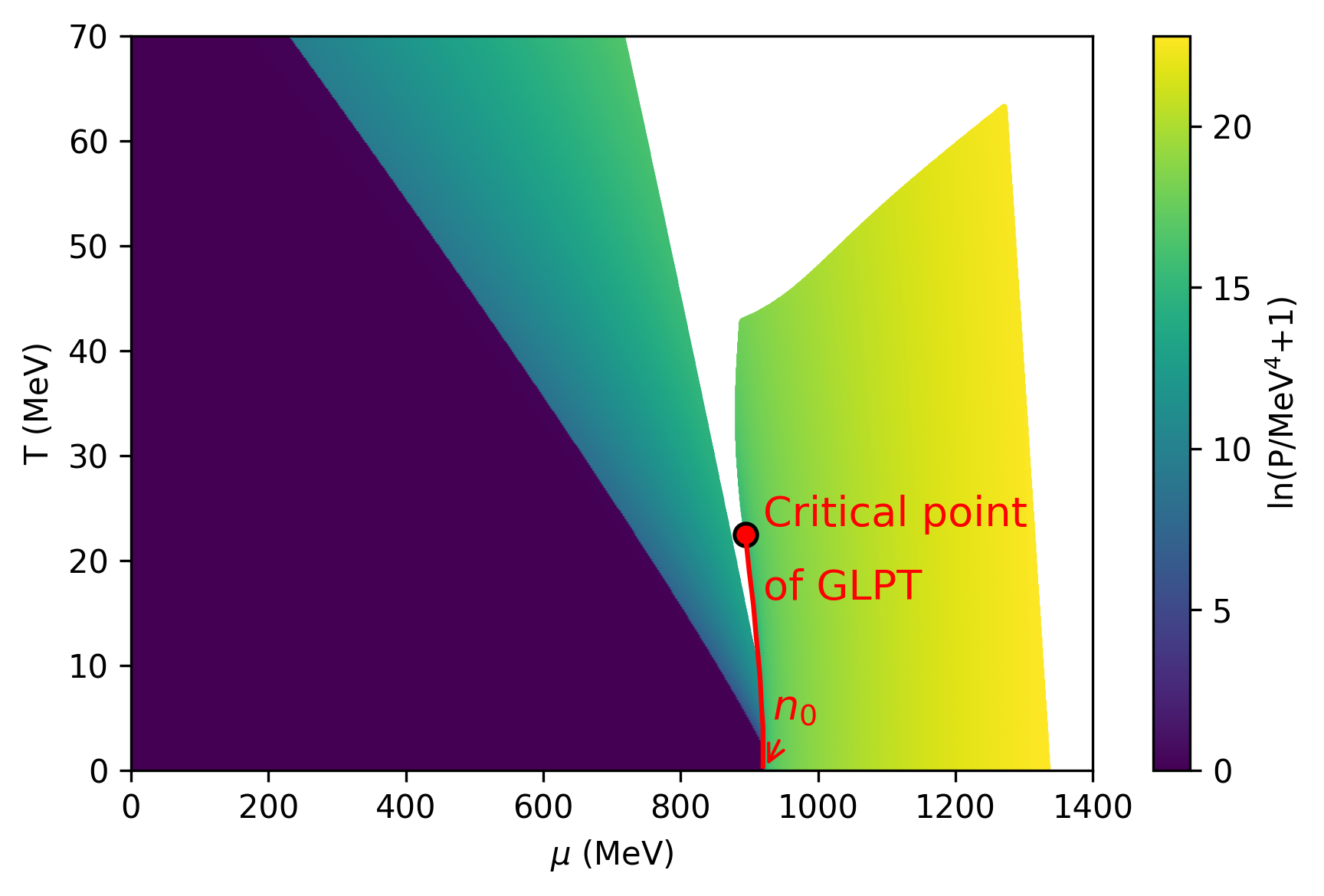}
  \caption{
    Pressure (\(-\Omega\)) for SNM in the valid regions of CSDC rules for the \(\mathcal{O}(k_{\mathrm{c}}^{12})\) parameter set in Table~\ref{tab:CCP} estimated by Eqs.~\eqref{eq:largemuD} and \eqref{eq:smallmuD} directly from the \(\mu_N^*\) and \(m_N^*\) values.
    The upper panel is for the \(30\%\) deviation region (\(\delta=30\%\)) and lower panel for the \(5\%\) deviation region (\(\delta=5\%\)).
    The thick blue regions correspond to the \(n=n_n+n_p\approx 0\) region.
    The regions with no color are out of the valid regions.
    }
  \label{fig:validregion}
\end{figure}

In order to keep the consistency of the \(k_{\mathrm{c}}\) expansion, by using Eqs.~\eqref{eq:largemuD} and \eqref{eq:smallmuD}, the valid regions of CSDC rules of \(\mathcal{O}(k_{\mathrm{c}}^{12})\) parameter sets in Table~\ref{tab:CCP} are given in Fig.~\ref{fig:validregion}. It can be seen that \(n\approx0\) regions shrink with temperature as discussed in the previous section, and the GLPT critical positions and nuclear saturation points are all within the valid regions of CSDC rules.

The LO [\(\mathcal{O}(k_{\mathrm{c}}^{4})\)] contribution of the CSDC rules is the free Fermi gas (FG) result, which does not have binding energy, and consequently no saturation point. The result of \(E/A\) for PNM shown in Fig.~\ref{fig:EPN_rho_PNM} is above the region given by leading-order chiral nuclear-force constraints~\cite{Drischler:2021kxf}.

The NLO [\((\mathcal{O}(k_{\mathrm{c}}^{6})\)] contribution is OBE level. It suppresses the \(E/A\) for PNM to make it slightly below the region given by the LO \(\chi\)NF, as the result of the emergence of a saturation point given by the cancellation between repulsive and attractive interactions. But the incompressibility at saturation density, \(K(n_0)\), is too large, around two times of the empirical value, as shown in Table~\ref{tab:RES}, resulting in a very stiff EOS, and makes \(E(n_a)\) is much smaller than empirical value at \(n_a=1.5 n_0\). Meanwhile, the symmetry energy slope, \(L(n_0)\), is also large.

When increasing the expansion to order \(\mathcal{O}(k_{\mathrm{c}}^{8})\), multimeson couplings such as \(\sigma^3\) and multimeson-nucleon couplings such as \(\sigma^2 NN\) start to contribute. These multimeson couplings will soften the EOS, lead to a much smaller \(K(n_0)\), a larger \(E/A\) at \(n_a=1.5 n_0\) and a smaller \(L(n_0)\), all closer to empirical values.
The \(E/A\) for PNM also aligns well with the LO \(\chi\)NF results, as shown in Fig.~\ref{fig:EPN_rho_PNM}.
But it makes \(T_{\rm c}\) higher than empirical value around \(20~\rm MeV\).

Let us continue to N\(^3\)LO, \(\mathcal{O}(k_{\mathrm{c}}^{10})\), where more multimeson and multimeson-nucleon couplings will involve. In this case, the larger \(T_{\rm c}\) is reduced and smaller \(L(n_0)\) value is improved, but \(K(n_0)\) becomes larger again with a small \(E(n_a)\).

Finally, at N$^4$LO, \(\mathcal{O}(k_{\mathrm{c}}^{12})\), all NM properties listed in Table~\ref{tab:RES} are well balanced and more consistent with empirical values, and the \(E/A\) for PNM is also within the region given by LO \(\chi\)NF~\cite{Drischler:2021kxf}.

Based on the above discussions, one can conclude that the NM properties at finite densities and finite temperatures can be captured by the CSDC rules with appropriate choice of the orders and parameters, and different order contributions have different physics involved and play different roles in determining the NM properties.

\subsection{Gas-liquid phase transition}
\label{GLPT}

\begin{figure}[tbh]
    \centering
    \includegraphics[width=0.45\textwidth]{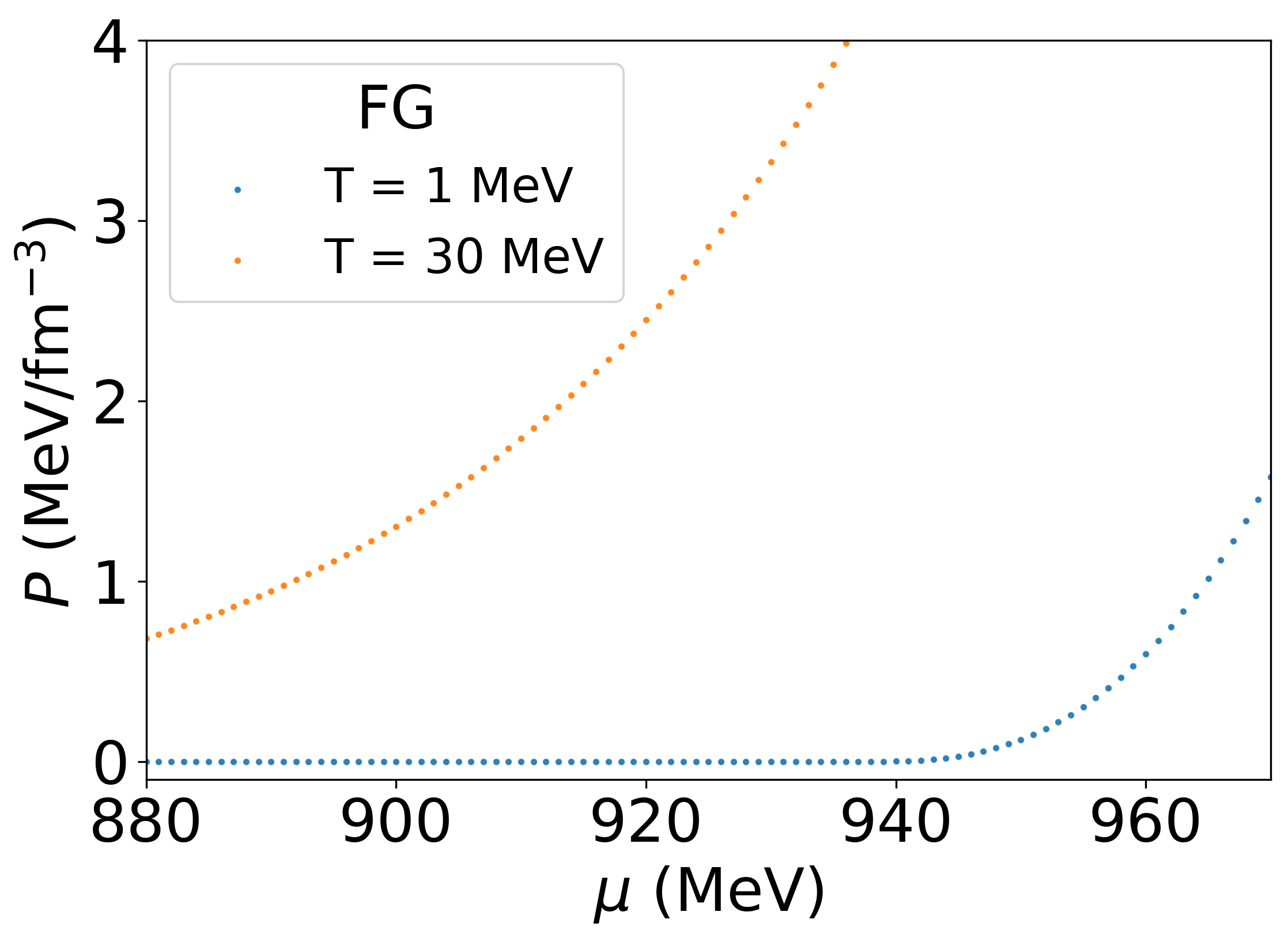}
    \includegraphics[width=0.48\textwidth]{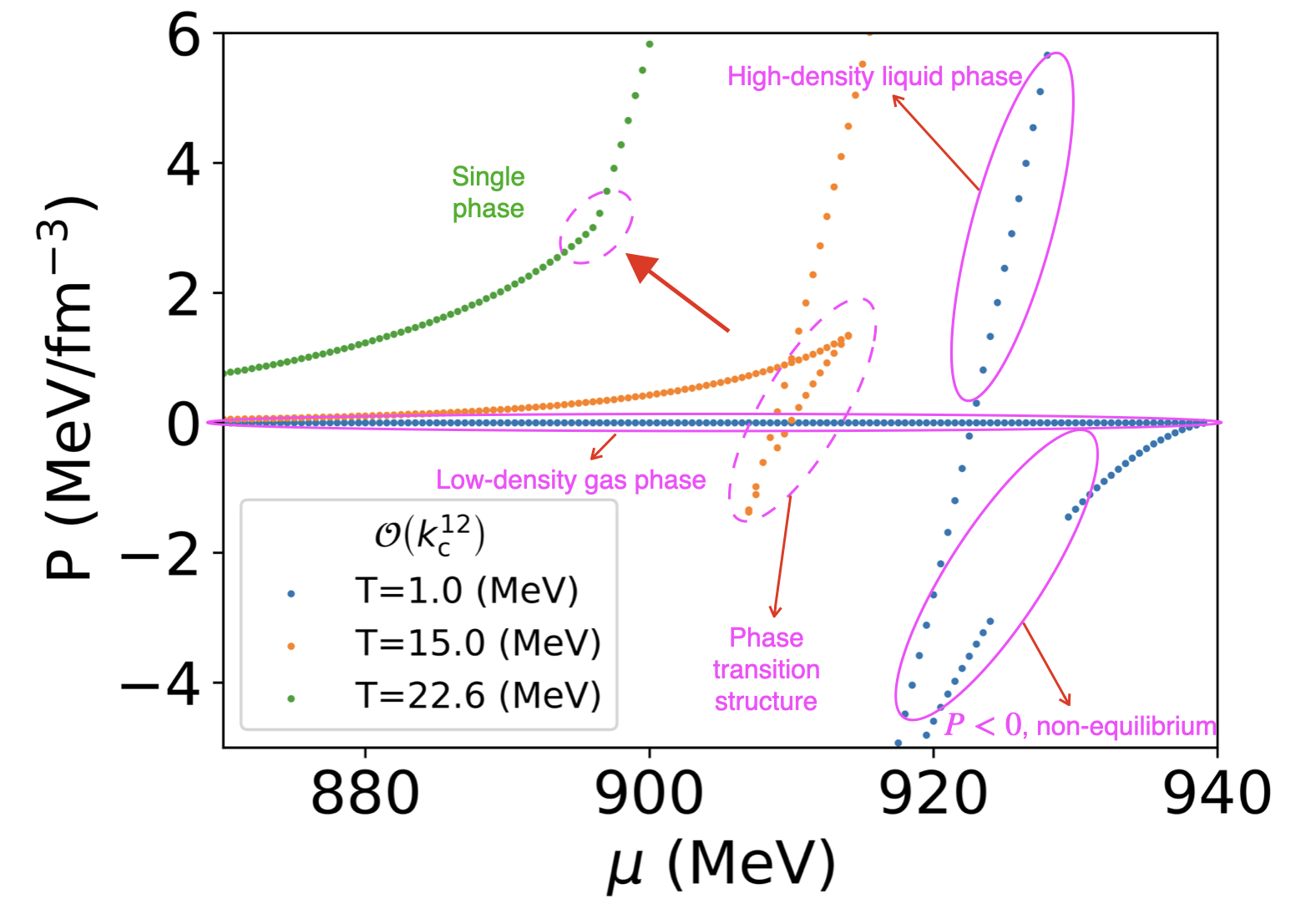}
    \includegraphics[width=0.45\textwidth]{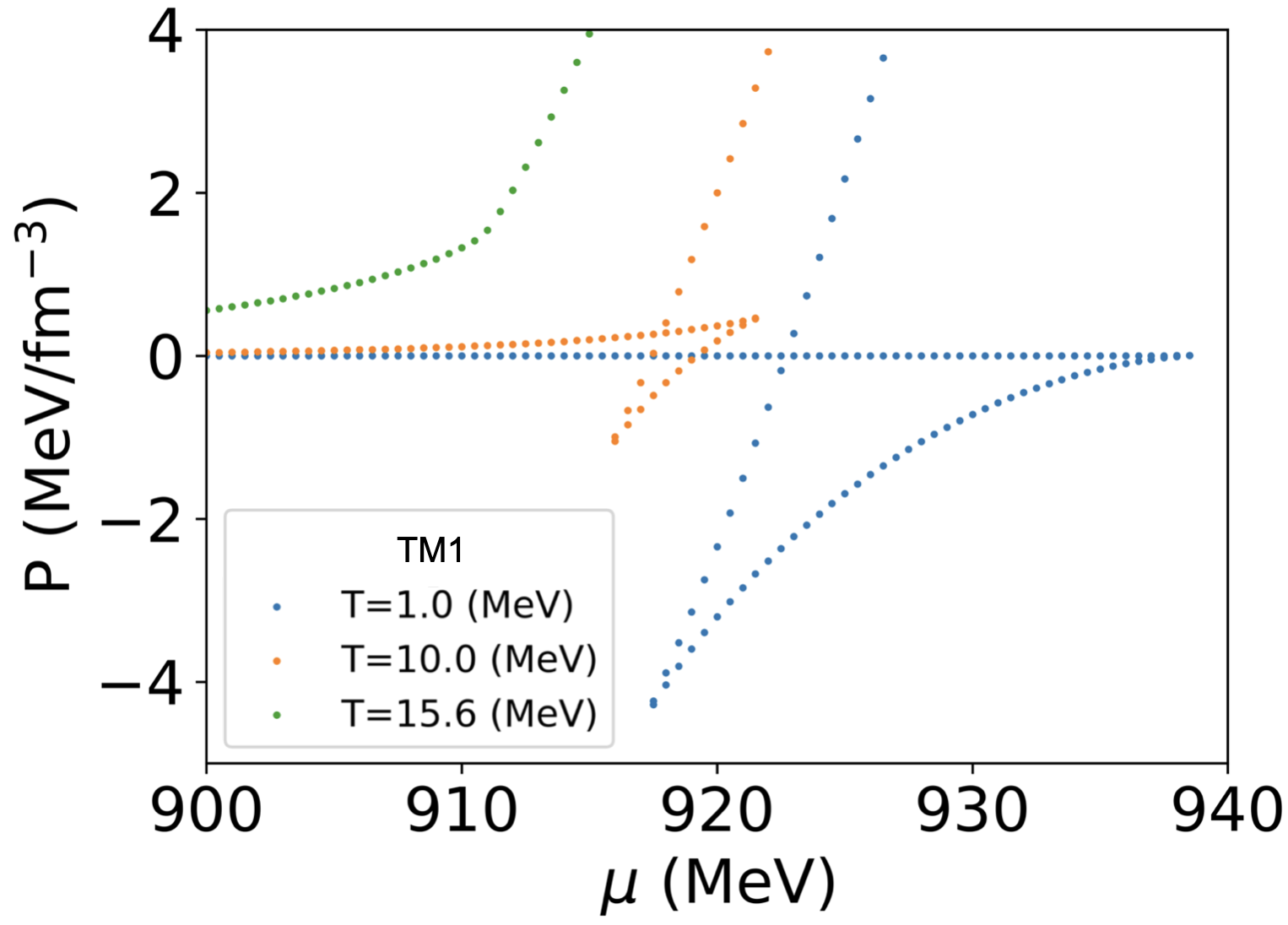}
    \caption{
        Pressure as a function of chemical potential for free Fermi gas (upper panel), \(\mathcal{O}(k_{\mathrm{c}}^{12})\) of CSDC framework (middle panel), and TM1 ~\cite{Sugahara:1993wz} lower panel).
    }
    \label{fig:P_mu_TM1}
\end{figure}

Let us now turn to more details on the GLPT of SNM. For clarity, we plot the pressure as a function of chemical potential for free Fermi gas (LO of CSDC framework), \(\mathcal{O}(k_{\mathrm{c}}^{12})\) of CSDC framework, and TM1~\cite{Sugahara:1993wz} in Fig.~\ref{fig:P_mu_TM1}.
  
The GLPT of SNM is from the binding energy of zero-temperature SNM, as a result of the competition between repulsive and attractive interactions, which is crucial to the formation of stable nuclei structures.
Then, the dependence of the pressure on chemical potential of SNM will be divided into three segments: a low-density gas phase, a high-density liquid phase, and an intermediate nonequilibrium phase. If the temperature is raised, these three segments will gradually merge into a single phase at the critical temperature, \(T_{\rm c}\), as illustrated in Fig.~\ref{fig:P_mu_TM1}: The binding energy structure is eliminated by temperature. 

From Fig.~\ref{fig:P_mu_TM1} one can see that there is no GLPT structure for the FG case, as expected. However, both the higher order CSDC and TM1 calculations exhibit the GLPT, and the segment of small \(\mu\) is called gas phase. Explicitly, it is found that \(T_{\rm c}\simeq 22.6~\rm MeV\) for \(\mathcal{O}(k_{\mathrm{c}}^{12})\) of CSDC, consistent with the estimation in Ref.~\cite{Karnaukhov:2003vp}, and \(T_{\rm c}\simeq 15.6~\rm MeV\) for the TM1 parameter set. Note that in the literature, the estimation of the GLPT is not convergent. It depends on the model used and how the data are extracted and ranges from several MeV to $20\pm3$~MeV~\cite{Karnaukhov:2003vp,Karnaukhov:2008be,Elliott:2013pna}.

In summary, the GLPT picture of SNM can be reproduced by CSDC framework, and aligns with the previous studies on NM EOS. This again confirms the rationality of our framework.

\subsection{Scale symmetry}

Next, let us turn to the behavior of scale symmetry at densities and temperatures. In our framework, the \(\sigma\) meson is considered as the NG boson of scale symmetry. It has been found that, the results of $bs$HLS saturates all the present constraints including the nuclear matter properties around saturation density, the mass-radius relations and the tilde deformability of NS~\cite{Ma:2018qkg}. In addition, it predicts the pseudoconformal structure of compact star matter which has not been observed before~\cite{Paeng:2017qvp,Ma:2018xjw,Ma:2018jze} and the peak of sound velocity in the intermediate density region, without resorting to any transitions from hadron to exotic configurations or introducing new degrees of freedom~\cite{Zhang:2024iye,Zhang:2024sju}.

\begin{figure}[tbh]
  \centering
      \includegraphics[width=0.45\linewidth]{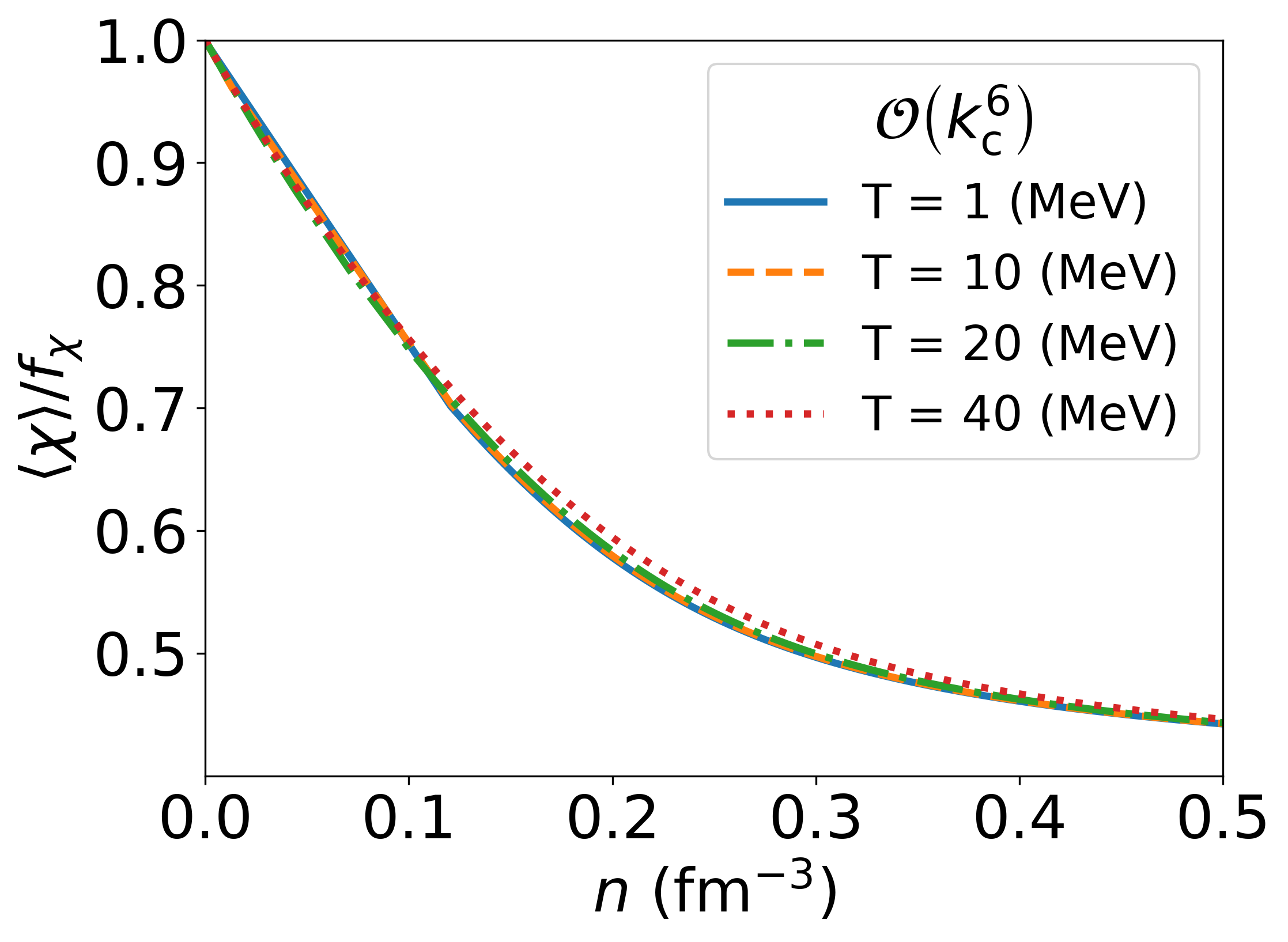}
      \includegraphics[width=0.45\linewidth]{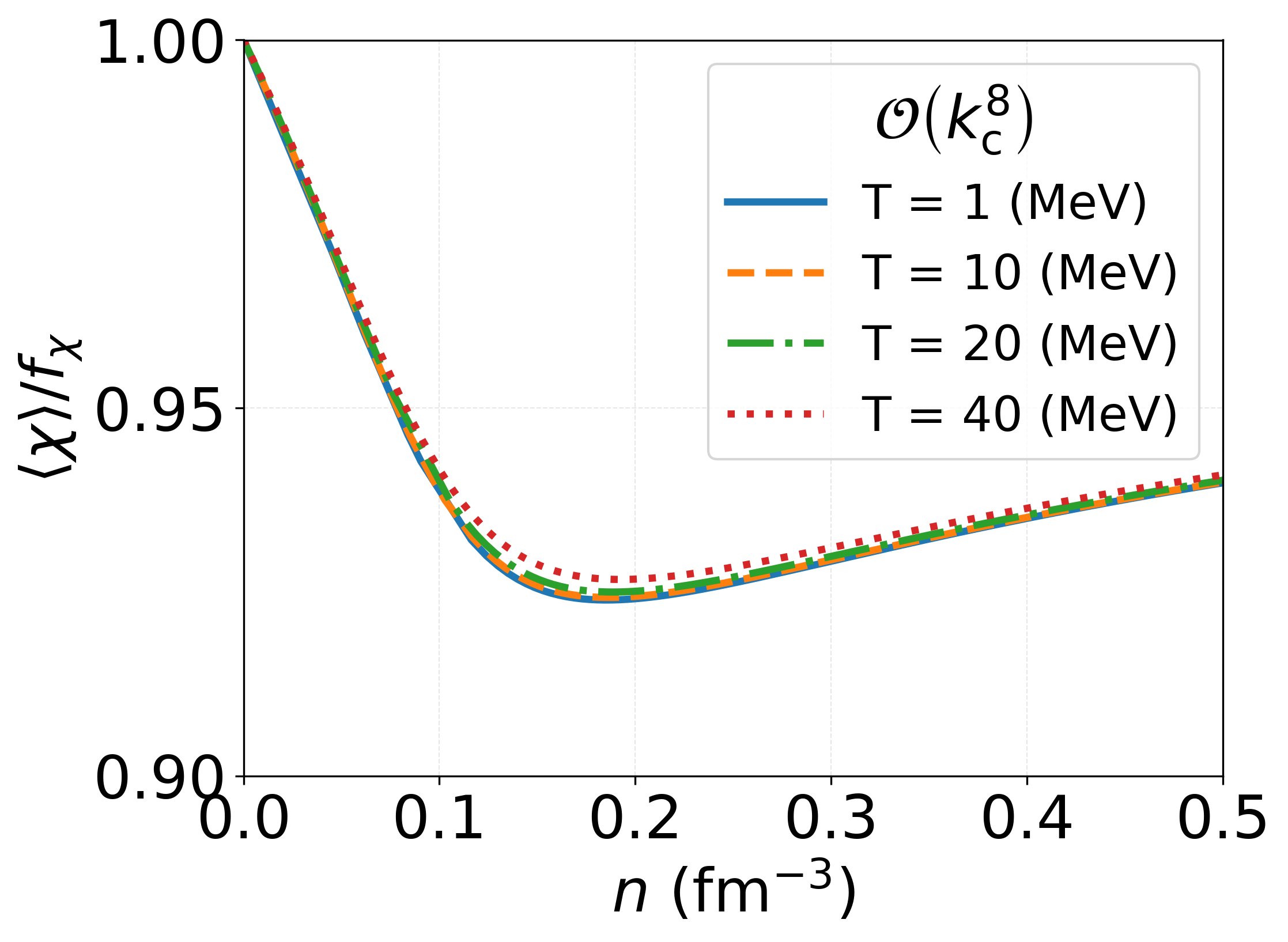}
      \includegraphics[width=0.45\linewidth]{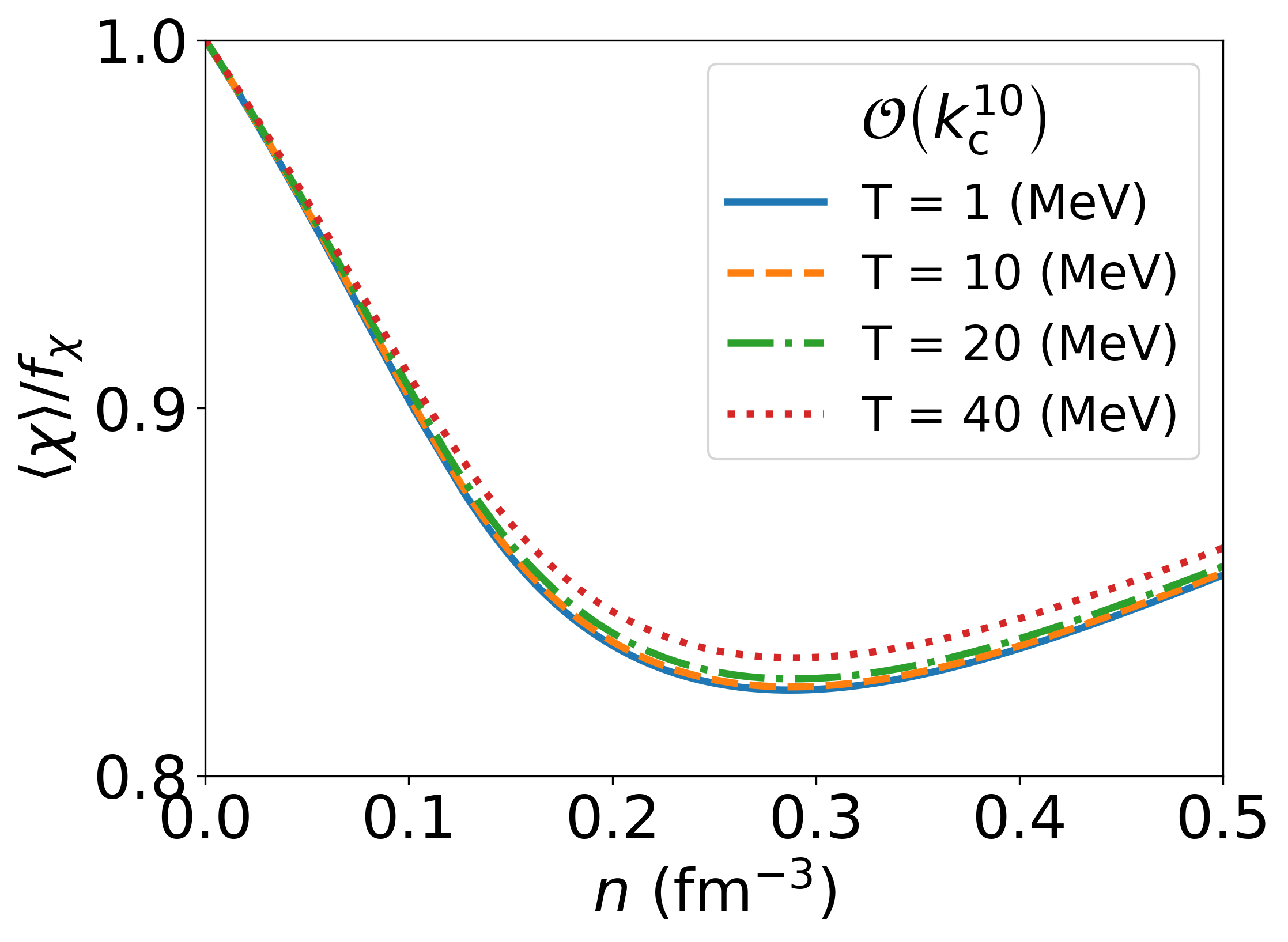}
      \includegraphics[width=0.45\linewidth]{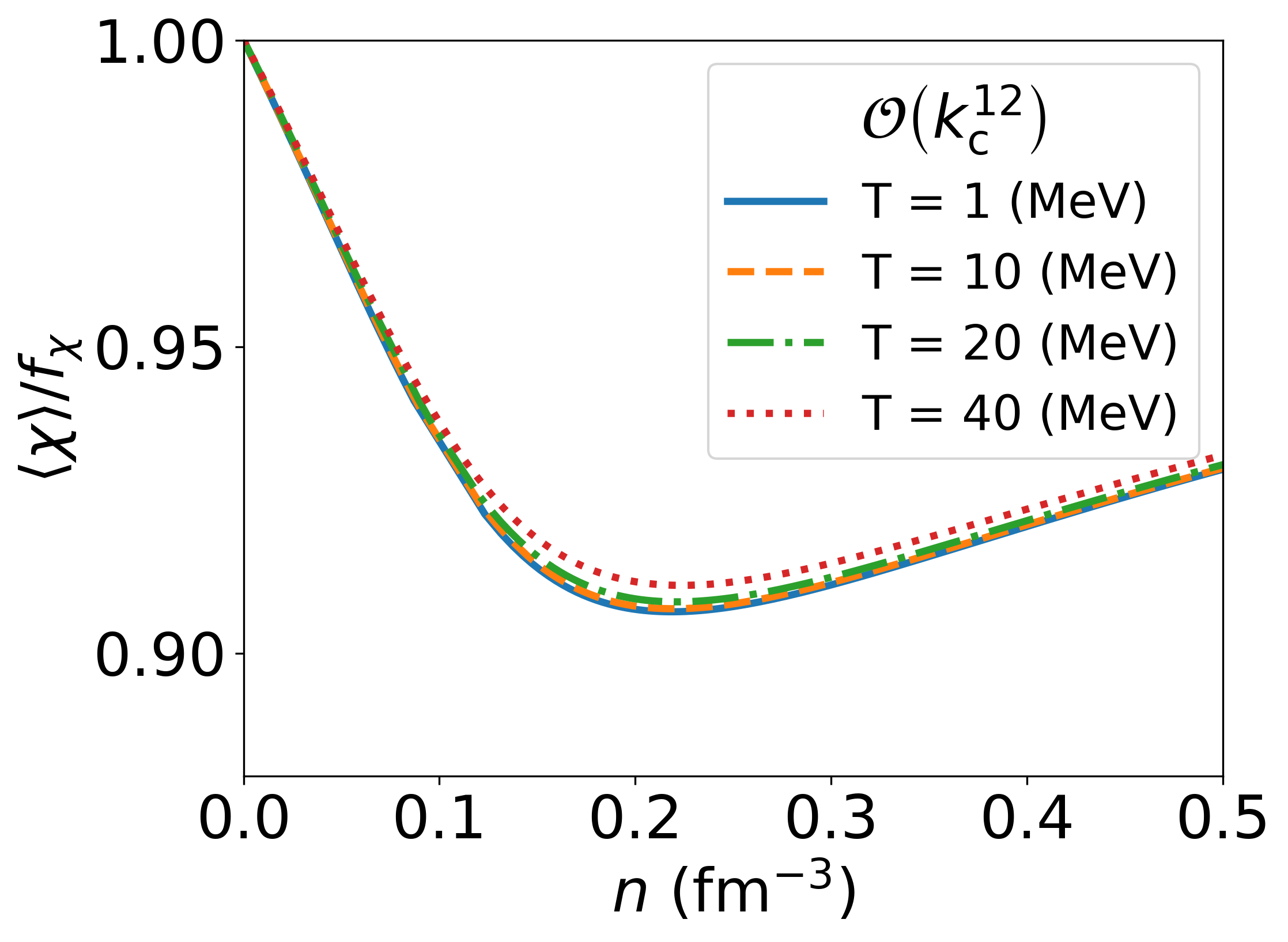}
      \includegraphics[width=0.45\linewidth]{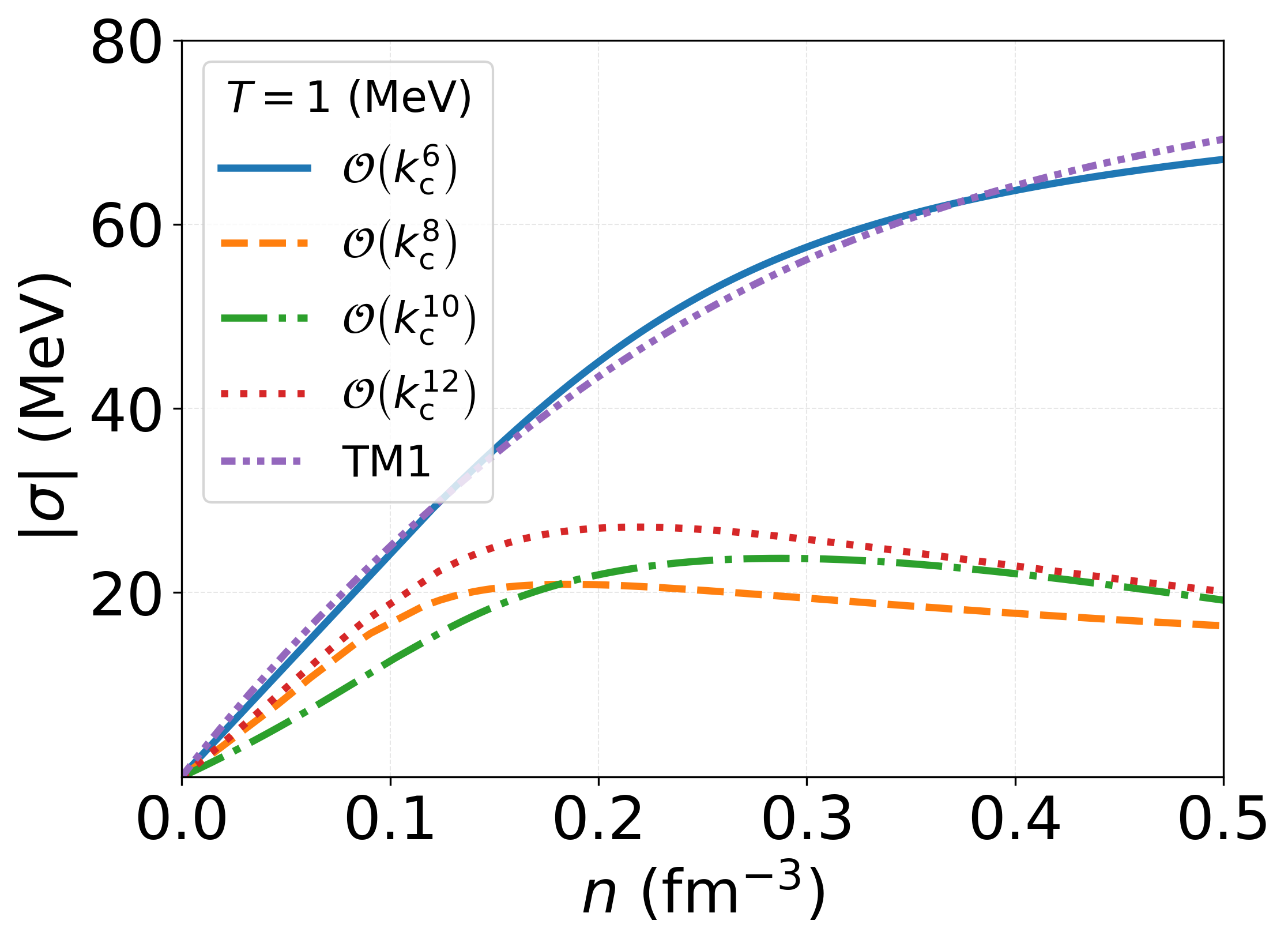}
    \caption{
      RMF \(\langle\chi\rangle\) as a function of nucleon number density at different temperatures for SNM.
      The results are calculated at $\mathcal{O}(k_{\mathrm{c}}^{6})$, $\mathcal{O}(k_{\mathrm{c}}^{8})$, $\mathcal{O}(k_{\mathrm{c}}^{10})$, and $\mathcal{O}(k_{\mathrm{c}}^{12})$.
      \(\sigma\) value for TM1 is calculated by the parameters in Ref.~\cite{Sugahara:1993wz} for comparison.
    }
    \label{fig:Combined_sigma_rho}
\end{figure}

The order parameter of scale symmetry is given by \(\langle\chi\rangle\), which can be estimated as \(\langle\chi\rangle\approx f_{\chi}e^{\langle\sigma\rangle/f_{\chi}}=f_{\chi}e^{\sigma/f_{\chi}}\) in the RMF approximation at densities and temperatures.
The RMF expectation of \(\chi\) as a function of density at different temperatures for each order up to \(\mathcal{O}(k_{\mathrm{c}}^{12})\) are shown in Fig.~\ref{fig:Combined_sigma_rho}.

It can be seen that the \(\langle\chi\rangle\) generally decreases with density at all CSDC orders at low densities, but those beyond OBE level will exhibit an increasing tendency at high densities. This is consistent with the findings in Refs.~\cite{Zhang:2024iye,Zhang:2024sju}, where such a kink structure resulting in a peak of speed sound.
The CSDC expansions confirm that the kink structure is from the unique couplings related to sigma meson (dilaton here), through the comparison between different orders and the TM1 set:
\begin{enumerate}
    \item At OBE level, one cannot distinguish the sigma meson is a chiral partner of pion or a dilaton;

    \item Beyond OBE level, compared to TM1, a dilaton sigma leads to more constrained relations between the couplings involving sigma meson from the matching to the QCD scale anomaly, see Eq.~\eqref{eq:morecoup}, resulting to the kink structure;

    \item For the changes with temperature, the \(\langle\chi\rangle\) does not shows much difference, and only slightly increase a little with temperature at different fixed densities. 
\end{enumerate}

\subsection{Sound velocity at finite density and temperature}

We finally study the SNM isothermal sound velocity \(C_{T}\) to see the effects of scale symmetry pattern.
The definitions of isothermal sound velocity is as follows~\cite{He:2022yrk},
\be
C_{T}^2&=&\frac{n}{T\left(\frac{\partial s}{\partial \mu}\right)_T+\mu\left(\frac{\partial n}{\partial \mu}\right)_T}\nonumber\ ,\\
\ee
which results into \(v_s^2={\rm d}P/{\rm d}\epsilon\) at \(T=0\).

\begin{figure}[tbh]
  \centering
    \includegraphics[width=0.45\linewidth]{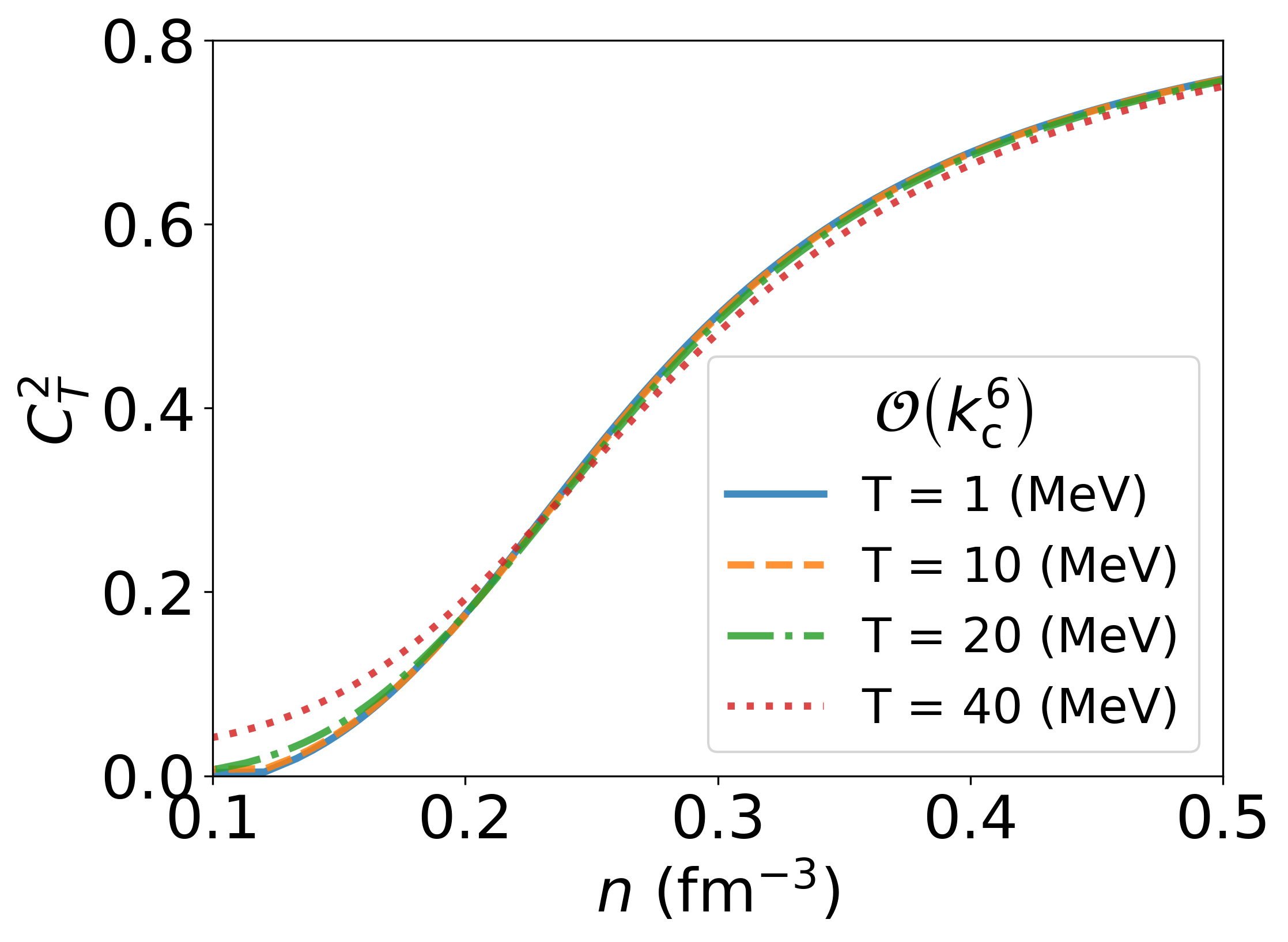}
    \includegraphics[width=0.45\linewidth]{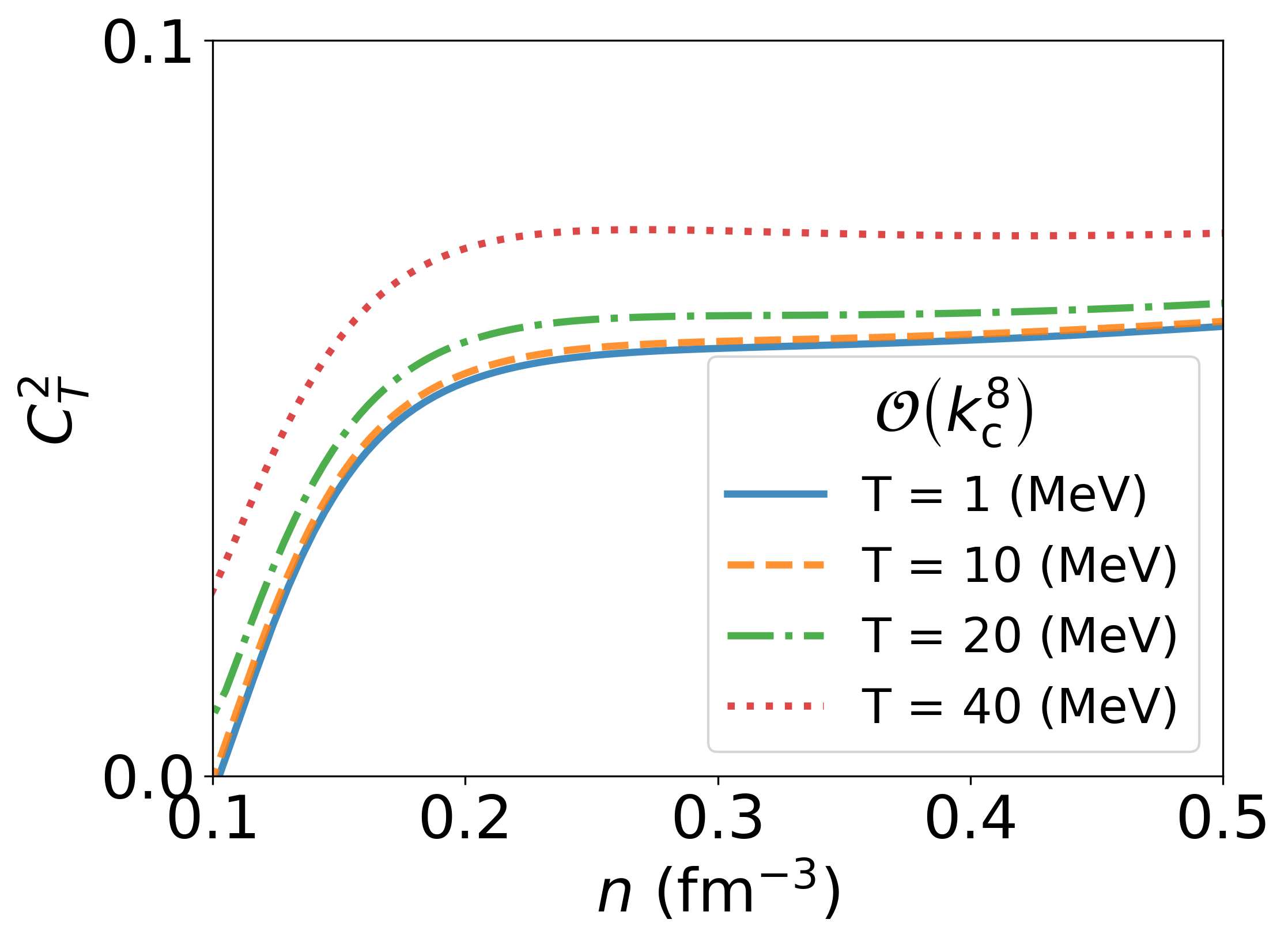}
    \includegraphics[width=0.45\linewidth]{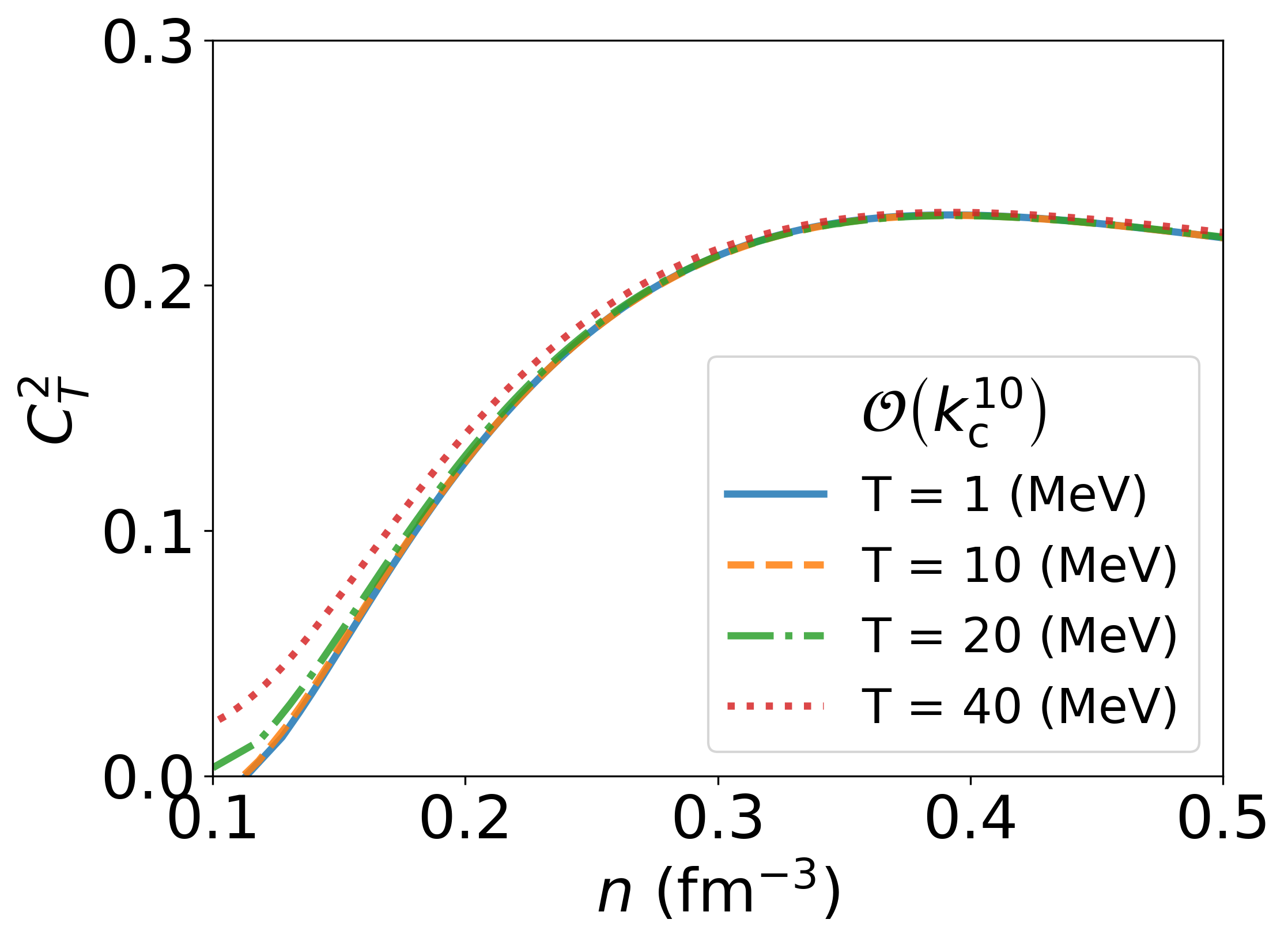}
    \includegraphics[width=0.45\linewidth]{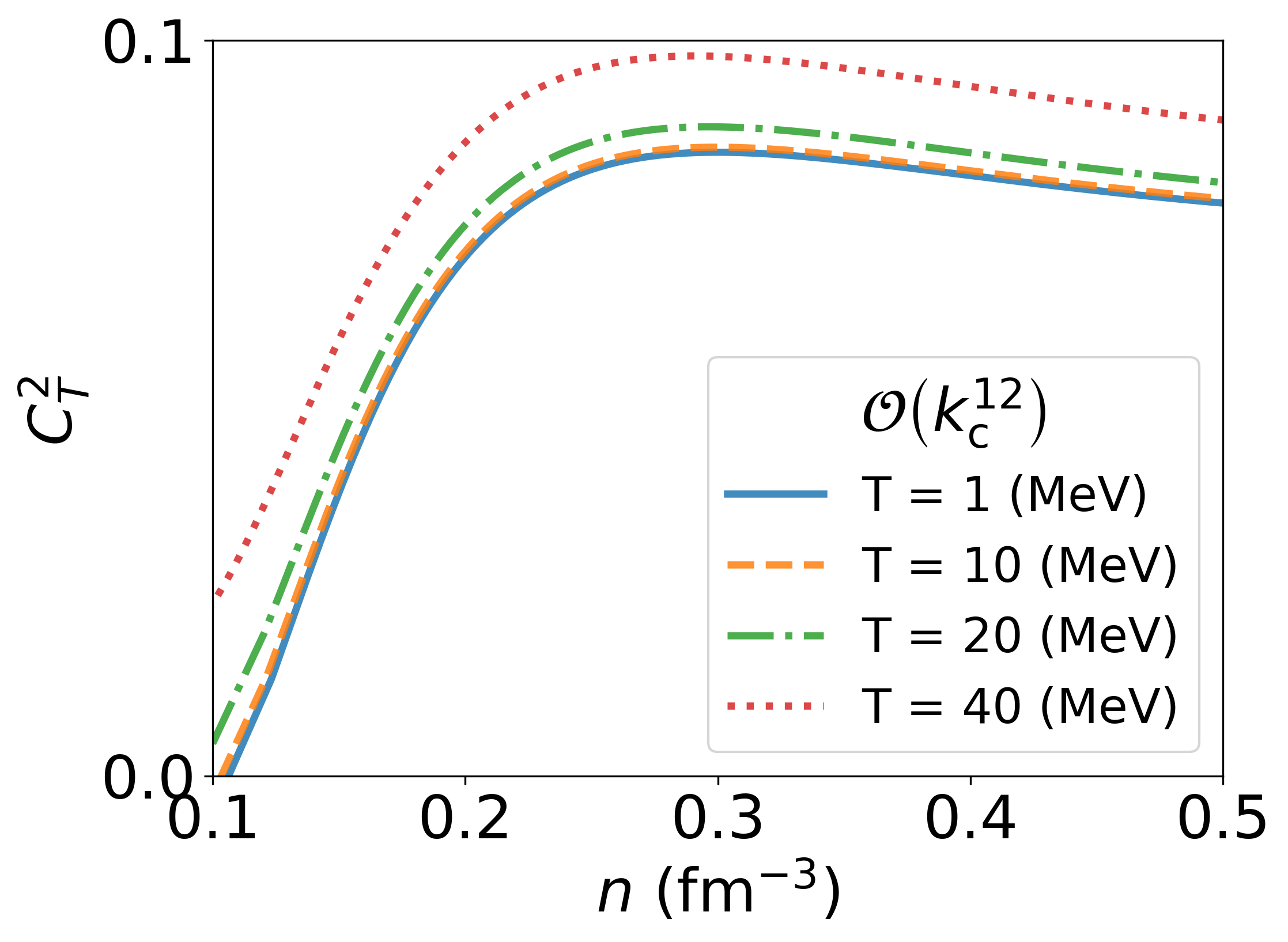}
    \includegraphics[width=0.45\linewidth]{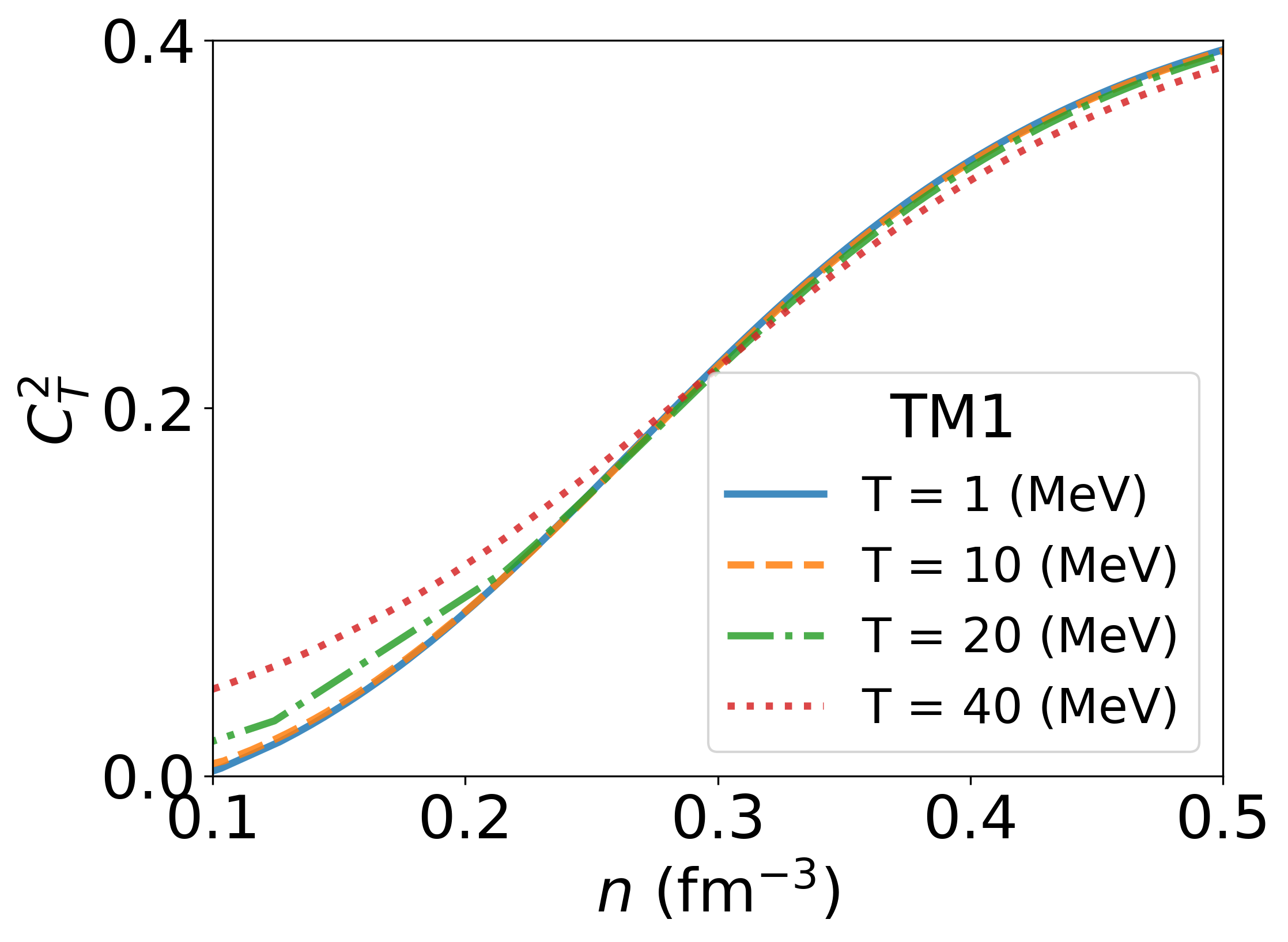}
  \caption{Isothermal sound speed, \(C_{T}^2\), as a function of nucleon number density at different temperatures up to $\mathcal{O}(k_{\mathrm{c}}^{12})$ and TM1~\cite{Sugahara:1993wz} for SNM.}
  \label{fig:Combined_Ct_rho}
\end{figure}

We can find from Fig.~\ref{fig:Combined_Ct_rho} that the OBE level and TM1 results exhibit a monotonic increase in \(C_{T}^2\) with density, while those beyond OBE level show a kink structure around \(2\sim3 n_0\), as the result of \(\sigma\) kink.
And the \(C_{T}^2\) increases with temperature at high densities for $bs$HLS, especially at \(\mathcal{O}(k_{\mathrm{c}}^{8})\) and \(\mathcal{O}(k_{\mathrm{c}}^{12})\), while the \(C_T^2\) from TM1 keeps almost unchanged.

There is one more point should be noted: The sound velocity, especially that for the beyond OBE level, is much suppressed than that from the TM1. This discrepancy was also found in Ref.~\cite{Ma:2025llw}, where a chiral effective model of QCD, baryonic extended linear sigma model, was applied to study the NS matter properties.
It was found that we can take care of both the physics at vacuum and the NS properties around/below saturation density simultaneously. But the inclusion of the physics at high-density regions, especially the NS core region, will lead to a contradiction:  Either the physics at vacuum or the NS properties around/below saturation density will be sacrificed to make the behaviors at high densities realistic in order to satisfy the observations of NS.

The reason for this contradiction lies in the different parameter and effective operator spaces, like $bs$HLS here and the extended linear sigma model, and the conventional Walecka-type RMF models, like the TM1. The former EFTs are constructed with respect to the symmetries of QCD, so that the vacuum physics, such as the pion decay constant \(f_{\pi}\) and mass spectra of hadrons, should be well described.
In this sense, the effective operators and their couplings are more constrained, e.g., \(\sigma^n\omega\omega\) terms in bsHLS are from the scale compensator to the \(\omega\) mass term, see Eq.~\eqref{eq:LM}, with the couplings related to \(m_{\omega}\) and \(f_{\chi}\). As a result, one can see in Table~\ref{tab:CCP} that there are much more multimeson couplings and their couplings are much larger than those in TM1. These larger and more multimeson couplings will make the behavior of EOS deviate more significantly from the OBE picture, which almost aligns with the Walecka-type models, see Fig.~\ref{fig:Combined_sigma_rho}.

Such a kind of contradiction may indicate the limitation of the RMF approximation, which ignore the meson exchange (Fock) terms, the full quantum effects of background fields, which will be significant at high densities. Another possibility is that the couplings/parameters should vary with medium (density and temperature) of the system in a wider region, through Brown-Rho scaling~\cite{Brown:1991kk} or density functional approach~\cite{Meng:2022nrx}.

\section{Summary and outlook}
\label{sec:sum}
 
In this paper, we established a systematic power-counting rule for the application of the chiral-scale EFT in NM studies, CSDC rules, at finite densities and temperatures.
Within this framework, free fermion gas is the LO contribution, consistent with the perturbation treatment of nucleon forces. NLO contribution is from the OBE interactions, while multimeson couplings involve at higher orders.

Then, we applied the CSDC rules in NM properties studies, and valid regions of the CSDC rules were provided. The zero temperature NM properties around \(n_0\) and the critical temperature of GLPT can be reproduced by \(\mathcal{O}(k_{\rm{c}}^{12})\), N\(^4\)LO of the CSDC rules with RMF approximation. And the EOS beyond the above regions is also consistent with the results of \(\chi\)NF. During this study, it is found that different orders capture the different physics of NM thermal properties.

Furthermore, the scale symmetry behavior was also investigated and found to be consistent with previous studies in Refs.~\cite{Zhang:2024iye,Zhang:2024sju}.
The scale symmetry restores at low densities, but rebreaks at high densities, resulting in a kink structure of \(\langle\chi\rangle\) and the sound velocity, \(C_T^2\), while temperature affects little.
This behavior stems from the consideration of \(\sigma\) as a dilaton, so that the couplings involving \(\sigma\) are constrained by matching the QCD scale anomaly.
Moreover, the sound velocity at high densities is much suppressed compared to the conventional Walecka type RMF models. This reminds us that the QCD symmetries may play an important role in determining the EOS for the fact that many effective operators, absent in Walecka-type models, cannot be neglected. In addition, we may go beyond RMF approximation and consider the scale dependence of the parameters in an analysis with rigorous consideration of QCD symmetries on a wide region of density, where vacuum physics, NM properties around \(n_0\), and NS structures are included.

In the near future, this study will be extended to study the proto NS, which depends on EOS of a wider region of density and temperature, and this will help us have a more insight on the CSDC rules.
Actually, the CSDC rules was established in the general MFA, not only under RMF approximation.
We plan to improve the study on NM properties by applying RHF scheme, and carry out a more rigorous numerical analysis on parameter space with Bayesian analysis.
We believe such a kind of research, using EFT of QCD to study NM properties at densities and temperatures, will help us understand strong interactions with such rich phenomena nowadays.

\acknowledgments

The authors would like to thank Qun Wang for helpful discussions.
The work of Y. M. is supported by Jiangsu Funding Program for Excellent Postdoctoral Talent under Grant Number 2025ZB516. Y.~L. M. is supported in part by the National Science Foundation of China (NSFC) under Grant No. 12547104, 12347103, and Gusu Talent Innovation Program under Grant No. ZXL2024363.

\section*{DATAAVAILABILITY}
The data that support the findings of this article are not publicly available upon publication because it is not technically feasible and/or the cost of preparing, depositing, and hosting the data would be prohibitive within the terms
of this research project. The data are available from the authors upon reasonable request.

\bibliography{refs}
\end{document}